\documentclass[aps, prx, reprint, longbibliography, nofootinbib,superscriptaddress, floatfix]{revtex4-2}

\usepackage{slashed}
\usepackage{verbatim}
\usepackage[T1]{fontenc}
\usepackage{mathbbol}
\usepackage[dvipsnames]{xcolor}
\usepackage{orcidlink}

\usepackage[normalem]{ulem}
\usepackage[english]{babel}
\usepackage{lipsum}
\usepackage{physics}
\usepackage{dcolumn}
\usepackage{tensor}
\usepackage{comment}
\usepackage{graphicx,color,overpic,mathtools}
\usepackage{amsthm,amsmath,amssymb,mathrsfs}
\usepackage{braket,bm,bbm,setspace}
\usepackage{cancel}
\usepackage{float}
\usepackage{xargs}
\definecolor{myred}{RGB}{179, 27, 27}
\usepackage{hyperref}
\hypersetup{
    colorlinks=true,
    linkcolor=myred, 
    citecolor=myred, 
    urlcolor=myred  
 }

\newcommand{\threeJ}[6]{\begin{pmatrix}#1 & #2 & #3\\ #4 & #5 & #6\end{pmatrix}}

\newcommand{\Amp}{\mathcal{A}}

\begin{document}

\title{Perturbative and nonlinear analyses of gravitational turbulence in spacetimes with stable light rings}

\author{Jaime Redondo-Yuste\orcidlink{0000-0003-3697-0319}}
\email[]{jaime.redondo.yuste@nbi.ku.dk}
\affiliation{Center of Gravity, Niels Bohr Institute, Blegdamsvej 17, 2100 Copenhagen, Denmark}

\author{Alejandro C\'ardenas-Avenda\~no\orcidlink{0000-0001-9528-1826}}
\affiliation{Computational Physics and Methods (CCS-2) \& Center for Nonlinear Studies (CNLS), \\Los Alamos National Laboratory, Los Alamos New Mexico 87545, USA}

\begin{abstract}
Some black hole mimickers, as well as black strings and other higher-dimensional spacetimes, exhibit stable light rings-regions where light or high-frequency gravitational waves can be trapped. In these regions, linear perturbations decay slowly, raising the possibility of nonlinear instability mechanisms. In this work, we study the cubic nonlinear wave equation as a proxy for Einstein's equations, using a four-dimensional model geometry that allows stable trapping. By employing a perturbative approach, and neglecting the backreaction onto the spacetime, we show that the nonlinear wave equation on the sphere with dissipative terms captures several features of the full nonlinear problem. This framework allows us to confirm a previous conjecture: all higher-order energy norms grow for arbitrarily small initial fluctuation amplitudes. Additionally, we analyze the system's mode spectrum at late times, revealing an inertial range dominated by a direct energy cascade. These findings further support the notion that spacetimes with stable light rings develop weak, high-frequency radiation hair, which will not generically lead to instabilities.
\end{abstract}

\maketitle

\section{Introduction}\label{sec:intro}

The final state conjecture posits that the end state of all gravitational processes are black holes (BHs), described by the Kerr metric, with some gravitational waves which ultimately radiate away. This conjecture can now be probed with unprecedented accuracy thanks to the combination of gravitational wave (GW) observations at different wavelengths through current ground based detectors, and future planned space based interferometers~\cite{LISA:2017pwj,LISA:2022kgy, Punturo:2010zz, ET:2019dnz, Reitze:2019iox}. Further insights can be obtained by observing the electromagnetic emission of accretion flow near supermassive black holes (BHs) with very long baseline interferometry~\cite{EventHorizonTelescope:2019dse,EventHorizonTelescope:2022wkp,Ayzenberg:2023hfw,EventHorizonTelescope:2024whi,Lupsasca:2024xhq}.

There are many theoretical proposals for compact objects~\footnote{We will refer generically to compact objects as those with compactness $\mathcal{C} = M/R \geq 1/3$, where $M$ and $R$ denote, respectively, the mass and the radius of the object or star.}, whose existence could violate the final state conjecture, with varying levels of development~\cite{Buoninfante:2024oxl}. Probably the first time this idea appears in the literature is as a ``geon,'' introduced by Wheeler~\cite{Wheeler:1955zz} searching for a self-gravitating configuration of the electromagnetic field. This is an example of a BH alternative that exists within a theory admitting BH solutions, such as General Relativity. In fact, although no stable geons were found, Wheeler's idea survived in the form of boson stars~\cite{Kaup:1968zz, Ruffini:1969qy,Liebling:2012fv}, gravitational condensates of a scalar field. Other proposals involve mechanisms that prevent gravitational collapse from ever forming event horizons (typically involving quantum gravity phenomenology), such as fuzzballs~\cite{Mathur:2005zp,Mathur:2024ify}, gravastars~\cite{Mazur:2004fk}, or anti--de Sitter (AdS) black shells~\cite{Danielsson:2017riq,Danielsson:2017pvl,Danielsson:2021ruf,Danielsson:2021ykm, Danielsson:2023onu, Giri:2024cks}. The term BH mimickers is sometimes used to refer generically to both classes of objects, and we refer the reader to~\cite{Cardoso:2016rao} for a comprehensive review on the topic. Here, we will focus on horizonless models, i.e., compact objects which do not have an event horizon, and are not singular either. We will briefly comment on the role of horizons later on. This class of BH mimickers was dubbed ``strong'' BH imposters in Ref.~\cite{Buoninfante:2024oxl}.

GW observations from compact binary mergers provide a means to test the BH hypothesis against various BH mimicker models~\cite{Berti:2015itd, Cardoso:2016oxy, Cardoso:2016ryw, LIGOScientific:2016lio, Yunes:2016jcc, Cardoso:2017cqb, Barack:2018yly, LIGOScientific:2021sio}. This can shed light toward a possible ultraviolet completion of General Relativity~\cite{Cardoso:2016oxy}, and help tackle the dark matter problem~\cite{Boehmer:2007um, Sikivie:2009qn}. Horizonless compact objects' most stark signature is the presence of GW echoes~\cite{Kokkotas:1995av,Ferrari:2000sr,Tominaga:1999iy,Tominaga:2000cs,Cardoso:2016oxy, Cardoso:2016rao, Price:2017cjr}. No evidence for this has been found in current observations~\cite{LIGOScientific:2021sio}. Gravitational-wave echo searches are typically based in parametrized models~\cite{Maggio:2019zyv, Maggio:2020jml, Maggio:2022nme}. However, different BH mimicker predict GW echoes arriving with vastly different time delays and amplitudes, making it challenging to rule out certain models only based on current observations.

An alternative approach is to investigate the generic dynamical properties of these objects. This implies questioning, e.g., how would such an object form dynamically, and its linear and nonlinear stability. Generically, horizonless objects that are also compact possess stable light rings (LRs). These are circular orbits where light rays can remain trapped indefinitely. Reference~\cite{Keir:2014oka} showed that the presence of stable LRs was associated to a slow decay of linear perturbations, and it was conjectured that this would lead to a nonlinear instability. The physical insight was that the trapping of high frequency modes for a very long time could trigger the formation of BHs, or the collapse or dispersion of the whole configuration. We refer to this as the LR instability conjecture. The goal of this work is to examine some of its fundamental aspects. 

Following the publication of~\cite{Keir:2014oka}, Ref.~\cite{Cardoso:2014sna} investigated the perturbative properties of BH mimickers with stable LRs. In particular, as it was already known for some stellar models~\cite{Ferrari:1984zz}, these objects have long-lived modes. Heuristic arguments favoring the presence of a nonlinear instability were discussed, as well. This problem has also been investigated from the numerical point of view, focusing on particular classes of mimickers that are solutions of well-posed equations of motion. Ref.~\cite{Guo:2024cts} showed that radial perturbations of scalarized, charged BHs with a stable LR, do not trigger any instability. In the context of boson stars, Ref.~\cite{Cunha:2022gde} claimed a generic instability for rotating boson stars with stable LRs, with a moderate timescale. The fate of the instability depends on the model under consideration, in some cases resulting in the collapse to a BH, while in others the star migrates toward a less compact configuration, without LRs. However, Ref.~\cite{Siemonsen:2024snb} was able to track the relaxation of a remnant boson star found after the merger of two such objects, where the remnant had similar properties as Ref.~\cite{Cunha:2022gde}, without any signs of an instability. Therefore, further numerical research and analytical insights~\cite{Zhong:2022jke} on these systems are paramount. 

A related problem is the stability of higher dimensional black strings and black rings, which also can possess stable LRs. This was first studied in detail in~\cite{Benomio:2018ivy}, proving also slow decay of linear perturbations, and conjecturing that this could also trigger a nonlinear instability. These problems exhibit similarities with the turbulent instability of pure anti--de Sitter (AdS) space. This spacetime, with reflective boundary conditions, is generically unstable against scalar perturbations~\cite{Bizon:2011gg, Buchel:2012uh, Buchel:2013uba, Buchel:2014xwa, Balasubramanian:2014cja, Balasubramanian:2015uua, Bizon:2011zz, Bizon:2014nhh, Bizon:2015pfa, Dias:2011ss, Dias:2012tq, Dias:2016ewl, Evnin:2021buq}. Although there are certain classes of initial conditions for which there is no such instability~\cite{Buchel:2013uba,Maliborski:2013ula, Balasubramanian:2014cja, Maliborski:2014rma,Balasubramanian:2015uua}, it is otherwise generic and needs only of two elements: (i) sufficient trapping of radiation, and (ii) a focusing mechanism, that localizes the radiation in a sufficiently small region close to the origin. The arguments in favor of the instability of BH mimickers with stable LRs invoke precisely this same two ingredients. Another system with similar features (rotating BHs in AdS) have also been shown numerically to be unstable against the development of high frequency hair~\cite{Figueras:2023ihz}. 

Our work is mainly motivated by a recent study \cite{Benomio:2024lev}, which investigates the dynamics of a defocusing cubic wave equation on a spacetime with stable trapping. By numerically evolving initial data localized within the stably trapped region, they identified a direct turbulent cascade in angular modes---a process in which higher and higher multipoles become excited, ultimately dominating the evolution. This mechanism arises due to the slow dispersion within the stable trapping region, where nonlinear interactions promote an energy transfer to increasingly small-scale structures. Crucially, based on their numerical results they conjectured that higher-order norms (or energies) of the solution can grow unbounded in time for generic initial data, in stark contrast with the decay properties typically seen when trapping is unstable (as in asymptotically flat black hole spacetimes like Schwarzschild). They further speculated that, even if the same turbulent behavior were to manifest for gravitational perturbations of compact objects or black string spacetimes in general relativity, it would not necessarily precipitate gravitational collapse or singularity formation.

In this work, we will prove the aforementioned conjecture regarding the unbounded growth of higher-order norms in the perturbative regime. Our setup is the same as in Ref.~\cite{Benomio:2024lev}, featuring in particular stable trapping, which underlies the slow dispersive properties driving the turbulent cascade. In this model, since backreaction on the spacetime is not included, the nonlinear behavior we study arises purely from the self-interaction of the scalar field. Establishing these results in a controlled, perturbative setting not only reinforces the numerical findings of Ref.~\cite{Benomio:2024lev}, but also clarifies how stable trapping can fundamentally alter the expected asymptotic behavior of nonlinear waves. Additionally we show that, unlike in pure AdS (see, for instance, Refs.~\cite{Bizon:2011gg, Buchel:2012uh,Balasubramanian:2014cja}), the confinement properties of this model are weak. Indeed, even at high frequencies, the radiation can be distributed in a number of overtones, which are long-lived and localized not at the LR itself, but within a ``light shell'' with non negligible thickness. Therefore, arguments based on the hoop conjecture to invoke the formation of BHs at the LR are bound to fail.

The structure of the paper is as follows. First, we revisit the model spacetime and nonlinear wave equation of Ref.~\cite{Benomio:2024lev} in Sec.~\ref{sec:set_up}. In Sec.~\ref{sec:linear} we solve the linear problem, studying the quasinormal mode spectrum of the system, and identify properties of the long-lived modes. We then examine the stability question from a mathematical point of view in Sec.~\ref{sec:norms}. We compute the growth rate of higher-order norms within the perturbative regime for a simplified configuration. Furthermore, we show how this calculation accurately estimates the growth of the norms obtained from the numerical evolution of the nonlinear wave equation in a spacetime with stable trapping. Lastly, we examine the development of a turbulent regime, and the end state of this process, in Sec.~\ref{sec:end_state}, and we conclude with a brief discussion of future prospects in Sec.\ref{sec:discussion}.
We relegate most of the technical details to Appendices~\ref{App:WKB} to~\ref{Appe:ConvergeTests}. In this work we use geometric units $G=1=c$, and choose the mostly plus signature of the metric $(-+++)$.

\section{The Setup: A simple geometry with a stable light ring}
\label{sec:set_up}

The LR instability is conjectured generically for many different hypothetical scenarios, based only on the existence of a closed, compact region with null trapping, such as a stable LR. However, the dynamics of different BH mimicker models are vastly different, since they involve different additional matter fields, or modified equations of motion~\cite{Cardoso:2019rvt}. We want to isolate the role of stable trapping at triggering a possible nonlinear instability when considering nonlinear propagating waves. Therefore, we want to exclude the possibility of coupling between different degrees of freedom (say, matter fields or GWs), as well as the effects of rotation (which may lead to the development of ergoregions, or superradiance, if there is absorption), or absorption of waves. This motivates the construction of a simple geometry which has a stable LR, and none of these other features. Because of this simplification, the resulting geometry is not a solution of Einstein equations. Nevertheless, the  nonlinear wave equation on this spacetime captures some of the features of Einstein equations~\cite{Benomio:2024lev}. In particular, we consider the same nonlinear wave equation studied in~\cite{Benomio:2024lev}, the defocusing cubic wave equation:
\begin{equation}\label{NLKG}
    \Box_g \Phi = \kappa \Phi^3 \, ,
\end{equation}
where $\kappa$ is a positive coupling constant that controls the strength of the nonlinearity, $\Phi:\mathcal{M}\to\mathbf{R}$ is a real field taking values in the manifold $\mathcal{M}$, endowed with the Lorentzian metric $g$. The operator $\Box_g$ is the d'Alembertian of the spherically symmetric metric
\begin{equation}
    g_{\mu\nu}dx^\mu dx^\nu = -f(r)dt^2+f^{-1}(r)dr^2+r^2d\Omega^2 \, ,
\end{equation}
where the radial function $f(r)$ will be specified below, and $d\Omega^2$ is the area element of the unit $2$-sphere. The choice of a cubic nonlinearity is motivated by the form of Einstein's equations in vacuum.  

While the spacetime we are considering is not a solution of vacuum Einstein's equations, we motivate the choice of the function $f$ by requiring that it has a pair of LRs at tunable locations. A simple functional choice that achieves this behavior is 
\begin{equation}\label{f}
    f = 1-\frac{8r^2}{R^2(4-\alpha^2)+\frac{(2+\alpha)^2}{R^2}r^4} \, ,
\end{equation}
where $R$ is the location of the stable LR, and $0<\alpha<1$ is a parameter that controls how ``deep'' is the cavity that forms at the stable LR. Notice that this spacetime has a vanishing ADM mass. This will further show that the features associated to the so called LR instability are exclusively due to the presence of the stable trapping region, and not sensitive to the low frequency structure of the geometry.

\begin{figure}
    \centering
    \includegraphics[width=\columnwidth]{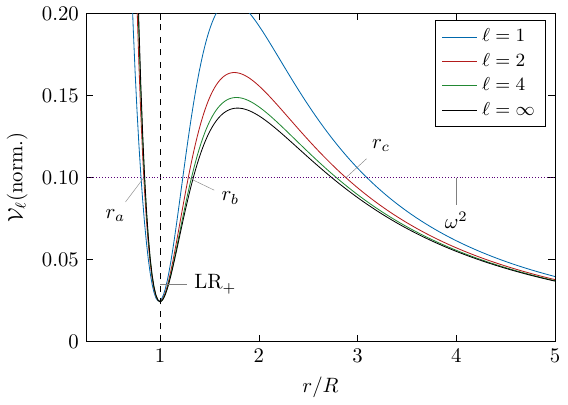}
    \caption{Normalized potential $\Tilde{\mathcal{V}}_\ell = \ell^{-1}(\ell+1)^{-1}\mathcal{V}_\ell $, defined in Eq.~\eqref{Potential_Definition}, with $\alpha=0.05$, for different values of the angular number $\ell$, including the eikonal limit (in black). The stable LR (denoted with a plus sign) is located at $r=R$. There is always another (unstable) LR, associated to the local maximum of the potential, at $r > R$. Additionally, we show for an arbitrary value of $\omega^2$ (purple dotted line), the three classical turning points $r_{a,b,c}$ which define the boundaries between the classically allowed and forbidden regions in the WKB approximation.}
    \label{fig:Potential}
\end{figure}

We can reduce the problem to a set of coupled ODEs by expanding the field in spherical harmonics. Additionally, to simplify the problem even further, and focus on the same scenario as~\cite{Benomio:2024lev}, we will restrict to axisymmetric modes. Thus, we can write
\begin{equation}
    \Phi = \sum_\ell \frac{\phi_\ell}{r}\mathcal{Y}_{\ell,0}(\theta) \, .
\end{equation}
Under this expansion, the nonlinear wave equation~\eqref{NLKG} becomes
\begin{equation}\label{NLKG_Modes}
    \mathcal{O}_\ell \phi_\ell = \kappa \sum_{\ell_1\ell_2\ell_3} \mathcal{I}^\ell_{\ell_1\ell_2\ell_3} \frac{f}{r^2}\phi_{\ell_1}\phi_{\ell_2}\phi_{\ell_3} \, ,
\end{equation}
where $\mathcal{I}$ is the overlap between the spherical harmonics, defined as 
\begin{equation}
    \mathcal{I}^\ell_{\ell_1\ell_2\ell_3} = \int_{4\pi}d\Omega \mathcal{Y}_{\ell_1,0}\mathcal{Y}_{\ell_2,0}\mathcal{Y}_{\ell_3,0}\mathcal{Y}_{\ell,0} \, , 
\end{equation}
which can be shown to be given by the following sum of products of Wigner $3j$-symbols
\begin{equation}\label{Mixing_Coefficients}
    \begin{aligned}
    \mathcal{I}_{\ell}^{\ell_1\ell_2\ell_3} =& \sum_{j=|\ell_1-\ell_2|}^{\ell_1+\ell_2} \frac{2j+1}{4\pi} \mathcal{C}^\ell_{\ell_1\ell_2\ell_3} \threeJ{\ell_1}{\ell_2}{j}{0}{0}{0}^2\threeJ{j}{\ell_3}{\ell}{0}{0}{0}^2 \, , \\ 
    \mathcal{C}^\ell_{\ell_1\ell_2\ell_3} =& \sqrt{(2\ell_1+1)(2\ell_2+1)(2\ell_3+1)(2\ell+1)} \,.
    \end{aligned}
\end{equation}
The system is, thus, governed by the following linear operator
\begin{equation}\label{Linear_Operator_Definition}
    \mathcal{O}_\ell = -\partial^2_t + f\partial_r\Bigl(f \partial_r\Bigr) - \mathcal{V}_\ell \, , 
\end{equation}
where the potential $\mathcal{V}_\ell$ is
\begin{equation}\label{Potential_Definition}
    \mathcal{V}_\ell = f\Bigl(\frac{\ell(\ell+1)}{r^2}+\frac{\partial_r f}{r}\Bigr) \, .
\end{equation}
The normalized potential for different values of $\ell$ is shown in Fig.~\ref{fig:Potential}. The geometry, by construction, possesses a stable LR at $r=R$, which leads to trapping of radiation, as well as an unstable LR located at $r \geq R$. 

Despite the nonlinear term, global solutions to Eq.~\eqref{NLKG} in the spacetime of a Schwarzschild black hole exist~\cite{bachelot1993equation, Nicolas1993}, and these arguments can be extended to our model spacetime. For this kind of nonlinearity, global existence follows from the conservation of energy, and the rate of decay of linear perturbations plays no role. The conserved energy norm is simply
\begin{equation}\label{NL_Energy_Norm}
    \mathbf{E}_{\rm nl}[\Phi](t) = \int_{\Sigma_t} d^3x \sqrt{-g}  \, T^{tt} \, , 
\end{equation}
where $\Sigma_t$ is a $t=\mathrm{const.}$ hypersurface, and $T^{tt}$ is a component of the stress-energy tensor, given by 
\begin{equation}
    T_{ab} = \partial_a\Phi\partial_b\Phi -\frac{1}{2}g_{ab}\Bigl(\partial_c\Phi\partial^c\Phi + \frac{\Phi^4}{2}\Bigr) \, .
\end{equation}
Energy conservation can be proven directly by taking the time derivative of~\eqref{NL_Energy_Norm}, applying the equations of motion and integrating by parts. Additionally, we can define high-order norms of order $k$, which can be seen as norms on the Sobolev space defined at each temporal slice, namely
\begin{equation}
\label{Sobolev_norm}
    \norm{D^{(k)}\Phi(t)}^2 = \int_{\Sigma_t} d^3x\gamma^{ij}\partial^{(k)}_i\Phi\partial^{(k)}_j\Phi \, ,
\end{equation}
where $\gamma_{ij}$ is the spatial metric on the $t=\mathrm{const}$ slice labeled by $\Sigma_t$. If we turn off the nonlinearity, i.e., when $\kappa=0$, these nonlinear norms are bound uniformly, and the local energy decays. More precisely, if $\Omega$ is a nonempty and bounded compact region, which does not change with time, and we denote by $\norm{D_\Omega^{(k)}\Phi(t)}^2$, the norm defined in Eq.~\eqref{Sobolev_norm}, but integrating only over the compact region, i.e., $\Sigma_t\cap\Omega$, we have~\cite{Benomio:2024lev}
\begin{equation}
    \norm{D^{(k)}_\Omega\Phi(t)}^2 \overset{\kappa=0}{\leq }\frac{C_{\Omega,k}}{\log(2+t)^2}\sum_{s=0}^{k+1}\norm{D^{(s)}\Phi(0)}^2 \, .
\end{equation}
However, as evidenced from the numerical simulations presented in Ref.~\cite{Benomio:2024lev}, it seems that this property no longer holds in the presence of the nonlinear term. This is one of the major attributes we analyze in the following sections. 

\section{Linear Perturbations}
\label{sec:linear}

Before analyzing the nonlinear problem, it is important to understand key aspects of the linearized problem. In this section, we will discuss the spectrum of the linearized problem, showing that there exists a family of long-lived modes associated with the stable LR. We characterize the frequencies and damping times of these modes in the eikonal regime by combining numerical methods that solve the linearized equations exactly with approximate methods valid in the eikonal regime. 

Let us solve the linear equation $\mathcal{O}_\ell \phi_\ell = 0$ subject to outgoing boundary conditions as $r\to\infty$, and look for solutions that are regular at the origin. This regularity condition can be seen as $\phi_\ell \sim r^{\ell+1}$ as $r\to 0$~\footnote{Irregular solutions behave as $\phi_\ell \sim r^{-\ell}$.}. Solutions satisfying these boundary conditions are quasinormal modes (QNMs). Using a shooting method we numerically solve the equation in the frequency domain by writing $\phi_\ell = e^{-i\omega_{\ell n} t} e_{\ell n}(r)$. 
\begin{equation}
    f\partial_r\Bigl(f\partial_r (e_{\ell n})\Bigr) + (\omega_{\ell n}^2-\mathcal{V}_\ell)e_{\ell n} = 0 \, ,
\end{equation}
subject to $e_{\ell n} \sim r^{\ell+1}$ as $r\to 0$, and $e_{\ell n} \sim e^{i\omega_{\ell n} r_\star}$ as $r\to\infty$, where the tortoise coordinate $r_\star$ is defined as the coordinate which satisfies $dr_\star = f^{-1} dr$. We implement these boundary conditions by expanding them in a power series, and solving for the asymptotic behavior of the equation up to a high enough order in powers of $r$ (close to the origin), or $1/r$ (at large distances). Then, we ``shoot'' for the QNM frequencies $\omega_{\ell n}$ by requiring that the solutions constructed from each boundary match smoothly at some intermediate radius~\cite{Chandrasekhar:1975zza, Berti:2009wx}. This unveils an infinite set of QNM frequencies and radial functions, labeled by the integer $n$, for each angular harmonic. The $n=0$ mode is the longest-lived, and therefore called the fundamental mode. Modes with $n>0$ have successively shorter damping times, and are referred to as the overtones of the system. 

\begin{figure}
    \centering
    \includegraphics[width=\columnwidth]{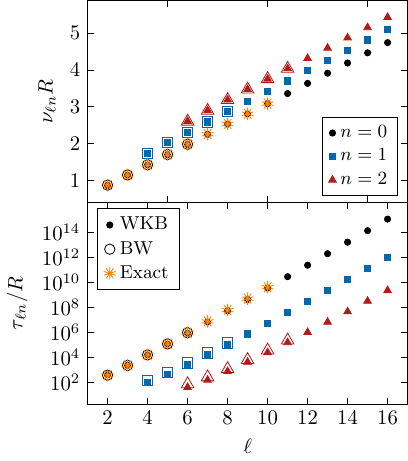}
    \caption{Oscillation frequency (top) and damping time (bottom) of the first few QNMs for the modeled spacetime with $\alpha = 0.166$. The exact values, computed with a direct integration, are shown as orange stars. Filled markers are computed using the WKB approximation, whereas the empty markers have been computed through the Breit-Wigner (BW) method. As expected, all three methods start showing good agreement for the higher harmonics $\ell \gtrsim 6$.}
    \label{fig:QNM_Freqs}
\end{figure}

This numerical method allows to obtain the fundamental QNM frequency for the first angular harmonics. The results can be visualized by writing $\omega_{\ell n} = \nu_{\ell n} - i/\tau_{\ell n}$, where $\nu_{\ell n}$ is the real part of the frequency of the QNM, and $\tau_{\ell n}$ its damping time. In particular, Fig.~\ref{fig:QNM_Freqs} shows how the damping time $\tau_{\ell n} R \sim e^{\gamma \ell}$ grows exponentially with $\ell$, with an exponential factor which can be computed directly from a Wentzel–Kramers–Brillouin (WKB) expansion (as shown in Appendix~\ref{App:WKB}), in the eikonal limit, as
\begin{equation}
    \gamma = \frac{\pi (f_{\rm LR_-}-\Omega^2)}{\sqrt{2 Q^{(2)}_{\rm LR_-}}} \, , 
\end{equation}
where $f_{\rm LR_-} = f(r_{\rm LR_-})$, and $Q^{(2)}_{\rm LR_-} = r_{\rm LR_-}^{-2}f^{\prime \prime}(r_{\rm LR_-})$, with $r_{\rm LR_-}$ the location of the unstable LR, and $\Omega^2 = f(R)/R^2$ the orbital frequency of null geodesics in the stable LR. 

These results confirm the intuition that modes with higher multipolar index $\ell$ are more efficiently trapped by the potential. Additionally, we confirm that the oscillation frequency scales linearly with $\ell$ as $\nu_{\ell} \approx \Omega \ell$. 

For the higher angular harmonics, Fig.~\ref{fig:QNM_Freqs} shows that these numerical solutions are in agreement with the WKB approximation, as well as a method based on Breit-Wigner resonances~\cite{Ferrari:1984zz, Pani:2009ss}. The details of these approximate methods are presented in Appendices \ref{App:WKB} and \ref{App:BWResonances}, respectively. Notice that the direct integration method requires higher numerical precision to accurately resolve the higher $\ell$ harmonics. Since this becomes quickly a computationally expensive task, we only compute up to $\ell = 10$.

These frequencies can also be extracted from the  numerical simulations of Ref.~\cite{Benomio:2024lev}, when using small initial data, to stay within the regime of validity of the linear theory. We highlight that for each $\ell$, there is a finite number of long-lived modes with different overtone index $n=0,\dots,N_\ell$. The highest possible overtone index for a long-lived mode $N_\ell$ can be estimated in the WKB approximation, as the largest integer such that  
\begin{equation}
    N_\ell +\frac{1}{2} \leq \frac{1}{\pi}\int_{r_a}^{r_{\rm LR_-}}\frac{dr}{f}\sqrt{\frac{f(r_{\rm LR_-})}{r^2_{\rm LR_-}}-\mathcal{V}_\ell} \, .
\end{equation}
Evaluating this expressions for the considered geometry, reveals that $N_\ell \lesssim \alpha \ell$ where $\alpha \in [1/3,1/2]$. Thus, in the high frequency regime, where all the damping times can be assumed to be much larger than any relevant dynamical timescale, there is not just one, but a very large number of long-lived modes. 

\section{Growth of the Norms}
\label{sec:norms}

Let us now examine the stability of the nonlinear problem, i.e., when $\kappa \neq 0$. While the nonlinear energy norm~\eqref{NL_Energy_Norm} is conserved, the behavior of the higher order Sobolev norms~\eqref{Sobolev_norm} does not have to. Note that one can also can define higher order energylike norms as 
\begin{equation}\label{Higher_Energy_Norm}
    \mathbf{E}^{(k)}_{\rm nl}[\Phi](t) = \int_{\Sigma_t} d^3x \sqrt{-g}  \, \, (T^{(k)})^{tt} \, , 
\end{equation}
with
\begin{equation}
    T^{(k)}_{ab} = \partial^{(k)}_a\Phi \partial^{(k)}_b\Phi - \frac{1}{2}g_{ab}\Bigl(g^{cd}\partial_c^{(k)}\Phi\partial_d^{(k)}\Phi + \frac{\Phi^2}{2}\Bigr) \, .
\end{equation}
These norms, as opposed to the Sobolev norms~\eqref{Sobolev_norm}, involve time derivatives, as well as the potential term. Ref.~\cite{Benomio:2024lev} presented numerical evidence that second order derivatives (related to the second order Sobolev norms) grow within their simulation time for sufficiently large initial data. It was conjectured that all norms with $k>1$ grow for arbitrarily small initial data. In this Section we address this question by computing the rate at which both Sobolev and energylike higher order norms grow, in the perturbative regime (therefore, for arbitrarily small initial data), in a simplified model. Lastly, we show, with additional numerical evidence, that the growth of the norms in the $3+1$ dimensional model follows a similar rate.

\subsection{Two-dimensional Model}

Let us begin by considering a simplified scenario. We consider the nonlinear wave equation on the circle $\mathbf{S}^1$. This system is not dissipative, and its spectrum of normal modes can be computed analytically. Concretely, let us solve the initial value problem 
\begin{equation}\label{1d_ivp}
    \begin{cases}
        -\partial^2_t \Phi + \partial^2_x \Phi &= \Phi^3 \, , \\
        \Phi(t=0,x) &= \phi_0(x) \, , \quad \partial_t\Phi(t=0,x)=0 \, ,
    \end{cases}
\end{equation}
where $x\in\mathbf{S}^1$, or equivalently, $\Phi(t,x)=\Phi(t,x+2\pi)$. We will make one more simplifying assumption: let us assume that the initial data is composed of a single mode. This means that we choose $\phi_0$ to be
\begin{equation}
    \phi_0 =  \epsilon\cos(n x) \, , \quad n\in\mathbb{N} \, .
\end{equation}
We will further assume that we are in the perturbative regime, so that $\epsilon \ll 1$. The solution, up to next to next to leading order (NNLO), can be written as 
\begin{equation}
    \Phi = \epsilon\phi^{(1)} + \epsilon^3 \phi^{(3)} + \epsilon^5 \phi^{(5)} +\mathcal{O}\left(\epsilon^{6}\right)\, , 
\end{equation}
where 
\begin{equation}
    \begin{aligned}
        \phi^{(1)} =& \cos(nx) \, , \\
        \phi^{(3)}=& a_1(t)\cos(nx)+a_3(t)\cos(3nx) \, , \\
        \phi^{(5)} =& b_1(t)\cos(nx)+b_3(t)\cos(3nx)+b_5(t)\cos(5nx) \, .
    \end{aligned}
\end{equation}
This results from expanding the higher-order equations into the normal modes of the circle with static initial data, represented by a cosine series. The nonlinearity is captured by the dependence of the modes on time. In particular, the time dependent coefficients satisfy the following equations of motion
\begin{equation}
    \begin{aligned}
        \ddot{a}_1+n^2a_1 =& -\frac{3}{4}\cos^3(nt) \, , \\
        \ddot{a}_3+(3n)^2a_3 =& -\frac{1}{4}\cos^3(nt) \, , \\
        \ddot{b}_1+n^2b_1 =& -\frac{3}{4}\cos^2(nt)(3a_1+a_3) \, , \\
        \ddot{b}_3+(3n)^2b_3 =&-\frac{3}{4}\cos^2(nt)(a_1+2a_3) \, , \\
        \ddot{b}_5+(5n)^2b_5 =&-\frac{3}{4}\cos^2(nt)a_3 \, .
    \end{aligned}
\end{equation}
The detailed solution is provided in the Appendix~\ref{App:NWE_Circle}. What matters for the current discussion is that the resonant terms on the right side of the above equations cause terms that grow secularly. To NLO we find linear growth in time, e.g., $a_1 \propto t\sin(nt)$, and to NNLO we observe terms that grow like $t^2$. This means that the NLO solution is only valid up to times when $t \leq \mathscr{O}(\epsilon^{-1})$, and the NNLO solution is only valid up to $t \leq \mathscr{O}(\epsilon^{-1/2})$. A two-timescale expansion~\cite{kevorkian2012multiple} would be necessary to find a solution valid for larger timescales.

We can now calculate the equivalents of the Sobolev and energylike higher-order norms in this situation. While the computation is straightforward, it is somewhat lengthy, and further details can be found in Appendix~\ref{App:NWE_Circle}. In this case, the Sobolev and energy higher-order norms are expressed as follows:
\begin{equation}
    \begin{aligned}
        \norm{D^{(k)}\Phi}^2 =& \int_{\mathbf{S}^1} dx \Bigl(\partial_x^{(k)}\Phi\Bigr)^2 \, , \\
        \mathbf{E}^{(k)}_{\rm nl}[\Phi](t) =& \int_{\mathbf{S}^1} dx \Bigl[\Bigl(\partial_t^{(k)}\Phi\Bigr)^2+\Bigl(\partial_x^{(k)}\Phi\Bigr)^2+\frac{\Phi^4}{4}\Bigr] \, ,
    \end{aligned}
\end{equation}
and the Sobolev norm, up to NNLO is given by 
\begin{equation}
    \begin{aligned}
        \norm{D^{(k)}\Phi}^2 =& \epsilon^2 \pi n^{2k}\Bigl[\cos^2(nt) + 2\epsilon^2\cos(nt)a_1(t) \\
        &+\epsilon^4 \Bigl(2\cos(nt)b_1(t) + a_1(t)^2 + 3^{2k}a_3(t)^2\Bigr)\Bigr] \, .
    \end{aligned}
\end{equation}
As we can see, the NNLO contribution has terms that grow like $t^2$ with slope $3^{2k}$ which depends on the order of the norm that we are considering. Let us be more explicit about this. If we define the rolling average over a period $T$ of the initial mode as 
\begin{equation}
    \langle f(t)\rangle_T \equiv \frac{n}{2\pi}\int_{T-\pi/n}^{T+\pi/n}dt f(t)  \, , 
\end{equation}
for any function $f(t)$, then, the rolling average of the Sobolev norm, after considering the time dependent mode coefficients, explicitly given in Appendix~\ref{App:NWE_Circle}, is 
\begin{equation}
    \begin{aligned}
        \langle\norm{D^{(k)}\Phi}^2\rangle_T =&  \frac{\pi}{2} \epsilon^2 n^{2k}\Biggl(1 + \frac{3\epsilon^2}{64^2}\Bigl[6\cos(2nT)-1\Bigr]\\
        &+\frac{72\epsilon^4 }{n^2}\Bigl[(360+3^{2k})T^2+\mathscr{O}(T) \Bigr]\Biggr)  \, .
    \end{aligned}
\end{equation}
It is now explicit that  $\langle\norm{D^{(k)}\Phi}^2\rangle_T \supset \epsilon^6 3^{2k}T^2$, whereas the other terms are either constants, or purely oscillatory terms. Thus, if we fix the period $T$ and the amplitude of the initial data $\epsilon$, we find that the higher-order norms have grown more relative to the lower-order norms. This is consistent with a direct energy cascade---if energy is flowing toward higher frequency modes, this must lead to faster growth of the higher energy norms since those weigh more heavily than the high-frequency content. 

A similar behavior occurs for the energylike higher-order norms. As shown in Appendix~\ref{App:NWE_Circle}, it leads to 
\begin{equation}\label{1d_nnlo_energy}
    \begin{aligned}
        \langle\mathbf{E}^{(k)}_{\rm nl}[\Phi]\rangle_T =& \pi n^{2k} \epsilon^2\Biggl(1  + 
        \frac{3\epsilon^2}{64}\Bigl[\frac{3}{n^{2k}}+n^{-2}\Bigl(6k-1\\
        &+3(1+(-1)^k)\cos(2nT)\Bigr)\Bigr]\\
        &+\epsilon^4 \Bigl[\frac{3^{2(k-1)}-1}{1024 n^2}T^2 + \mathscr{O}\Bigl(T, T^2\cos(nT)\Bigr) \Bigr]\Biggr) \, .
    \end{aligned}
\end{equation}
By setting $k=1$ we recover a constant value, as expected, 
\begin{equation}
    \label{Eq:Growepsilon4}
    \langle\mathbf{E}^{(1)}_{\rm nl}[\Phi]\rangle_T= \pi n^2 \epsilon^2  + \frac{3\epsilon^4}{8} = \mathrm{constant} \,.
\end{equation}
We have also checked that the NNLO term vanishes exactly, even before taking the average. However, for the higher-order norms, we observe the same behavior for both the Sobolev and energylike norms, i.e., the norms grow as $\epsilon^6 c_k T^2$, where $c_k$ is some monotone function of $k$, which depends on the precise definition of the norm. Note that the energylike norms must have $c_{k=1}=0$, whereas this restriction does not apply to the Sobolev norms, which can grow even at first order. 

\subsection{Four-dimensional model}

Building on our analysis of the two-dimensional model, we now return to the original four-dimensional setup to apply the insights gained and further explore its behavior. Unlike the simpler nonlinear wave equation on a circle, the four-dimensional system exhibits several key differences. First, it is dissipative: because the scalar field can escape to infinity rather than remain perfectly trapped at the stable LR, the mode frequencies are complex. Second, in the $2$-dimensional case, frequencies of normal modes are commensurate. In contrast, for the four-dimensional setup, the sum of two (or more) QNM frequencies does not generically match the frequency of any other QNM, so the system is not at an exact resonance. 

Although notable differences exist between the previous model (the nonlinear wave on the circle) and the physical scenario of interest, we can still examine whether the signature of growing higher-order norms emerges here. A key similarity is evident: if we excite the system with the \(\ell = L, n=0\) mode, modes with \(\ell = 3L, 5L, \dots\) subsequently appear, hinting at a direct cascade of energy.

Estimating the full higher-order Sobolev or energy norms from numerical solutions is a complicated task. Extracting reliable estimates of higher derivatives--and thus higher-order norms--from numerical solutions can quickly become cumbersome, as each (numerical) differentiation compounds the approximation error. One may also require (numerical) interpolation for the (numerical) integration over the domain. Nonetheless, building on the results of~\cite{Benomio:2024lev}, we know that angular derivatives at the LR can grow rapidly for large initial data, which indicates that the angular terms in both Sobolev and energy norms might increase as well. 

We solve the nonlinear wave equation~\eqref{NLKG}, for axisymmetric initial data, using the numerical method described in Ref.~\cite{Benomio:2024lev}. We prescribe initial data localized around the stable LR. Since this is a trapping region, the field is confined at the LR, and its radial structure remains largely unchanged. Therefore, we can confidently estimate the total energy norm by extracting only the norm at the LR itself. We define, assuming additionally an axisymmetric configuration, the norm at the LR as
\begin{equation}
    \norm{D^{(k)}\Phi(t)}^2_{\rm LR} = \int_{r=R} \sin\theta d\theta d\phi \Bigl(|\partial^{(k)}_\theta \Phi|^2 
    \Bigr) \, ,
\end{equation}
and conjecture that all the norms scale in a similar manner, i.e., $\mathbf{E}^{(k)}_{\rm nl}[\Phi(t)] \sim \norm{D^{(k)}\Phi(t)}^2 \sim \norm{D^{(k)}\Phi(t)}^2_{\rm LR}$. Now, if the field at the LR is expanded in spherical harmonics (assuming axisymmetry)
\begin{equation}
    \Phi = \sum_\ell c_\ell \mathcal{Y}_{\ell,0}(\theta) \, ,
\end{equation}
then one can compute the localized norm at the LR, up to a positive numerical factor $b$, as
\begin{equation}
    \norm{D^{(k)}\Phi(t)}^2_{\rm LR} = 2\pi b \sum_\ell \ell^{2k} c^2_\ell  \, .
\end{equation}
Without loss of generality, we will now set $b=1$. The challenging part about extracting the behavior of these norms from numerical simulations is that, due to numerical errors, we only have access to the lower order multipoles, roughly, $\ell \lesssim 16$. In particular, for the values of $\alpha,R$ that we choose for numerical purposes, $\alpha = 0.166$ and $R = 0.229$ (chosen to agree with~\cite{Benomio:2024lev}, which make it simpler for the potential well to be numerically resolvable after compactifying the domain), the damping time of the $\ell=2$ mode is only $\tau_{20}/R = 3.8\times 10^{2}$ (see Fig.~\ref{fig:QNM_Freqs}). If we excite initially the $\ell = 2$ mode, we will only see the norm growth on timescales shorter than this decay timescale, $t \ll \tau_{20}$. From the $2$-dimensional model, we may expect that the norm grows as $\epsilon^4 t^2$, as shown in Eq.~\eqref{1d_nnlo_energy}. The high power of the amplitude $\epsilon$ in this scaling makes it challenging to observe the norm growth for small initial data, even for the higher order norms. 

\begin{figure}
    \centering
    \includegraphics[width=\columnwidth]{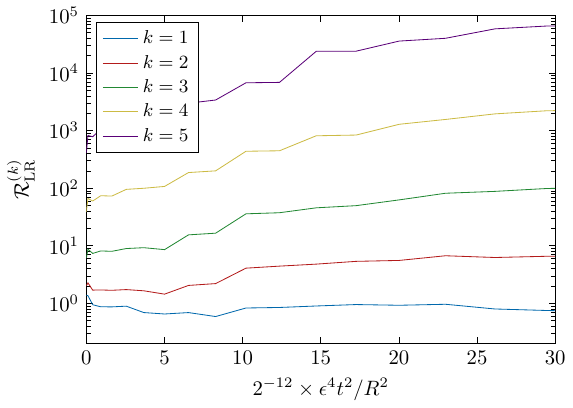}
    \caption{Ratio of the $k$-th order norm with respect to its initial value for different values of $k$, as indicated in the legend. The initial values are defined as in Eq.~\eqref{energy_ratio}, at different times, with respect to the rescaled time $\propto \epsilon^4 t^2/R^2$, a factor based on Eq.~\eqref{1d_nnlo_energy}. We choose a relatively large amplitude $\epsilon = 1.5$, in principle outside of the regime of validity of perturbation theory, where the initial data excites only the $\ell=2$ mode. We set $\alpha=0.166$ and $R=0.229$ in Eq.~\eqref{f}.}
    \label{fig:3d_norm_growth}
\end{figure}

We examine the norm ratio
\begin{equation}\label{energy_ratio}
    \mathcal{R}^{(k)}_{\rm LR} \equiv \frac{\norm{D^{(k)}\Phi(t)}^2_{\rm LR}}{\norm{D^{(k)}\Phi(0)}^2_{\rm LR}} \, ,
\end{equation}
in Fig.~\ref{fig:3d_norm_growth}, for different values of $k$, and an initial amplitude of $\epsilon =1.5$ (which sets us away from the perturbative regime). Figure~\ref{fig:3d_norm_growth} shows how this norm ratio grows in an approximately linear fashion with $\epsilon^4 t^2$ (mind this scaling for the horizontal axes), during timescales shorter than the linear decay time of the fundamental mode. However, since the amplitude of the initial data may not be considered small, the agreement between this Figure and Eq.~\eqref{1d_nnlo_energy} is remarkable. Moreover, we can observe how the higher order norms (larger values of $k$) grow faster than the lower values of $k$. This is the same feature present already in the two-dimensional model. The ratio does not initially start at $1$ for the higher-order norms because we average over multiple time points, and the rapid energy exchange between modes makes this transient phenomenon difficult to resolve.

These results provide analytical evidence that in the presence of stable trapping, nonlinear dynamics lead to a direct energy cascade, which in turn results in growing higher-order norms.  In particular, higher-order norms grow faster than lower-order norms, and the growth rate seems to be polynomial in time. The norms do not diverge in finite time--it has been already established the global existence of solutions for this equation. This implies, in particular, that for any initial data $\epsilon$ and finite time $T$, we can always find an integer $K(\epsilon,T)$ such that norms with order $k \geq K$ have grown more than a certain factor of their initial value, regardless of how small this initial data is.

\section{The End State of the Gravitational Turbulent Cascade}
\label{sec:end_state}

The previous discussion throws some light to answer a question posed in previous work~\cite{Benomio:2024lev} related to the behavior of higher-order norms in nonlinear systems in the presence of stable trapping. As we have found, all these norms grow, regardless of how small the initial data is. These higher order norms do not have a clear physical interpretation. In fact, the only norm with a well-defined physical meaning is the energy norm~\eqref{NL_Energy_Norm}, which is actually conserved. The growth of these norms is interpreted as a consequence of the direct energy cascade. However, major significant questions remain unanswered. How does this cascade develop, and what are its properties? What is the end state of this dynamic process? In this section, we set ourselves to address these questions. 

\subsection{Two-dimensional Model}

As done in the previous section, let us start with the lower-dimensional model, i.e., the nonlinear cubic wave equation in the circle. We evolve numerically the system of Eq.~\eqref{1d_ivp} with a pseudospectral strategy. We use the \texttt{ApproxFun.jl} package~\cite{olver2013fast} to compute the nonlinear term in real space, and then evolve the Fourier representation of the equation in time using an explicit, fourth order Runge-Kutta method. After each step, we dealias the latter half of the Fourier modes to filter out contributions that can numerically contaminate the lower-frequency modes~\cite{boyd2001chebyshev}. A convergence test for our numerical solutions is shown in Appendix~\ref{Appe:ConvergeTests}. 

\begin{figure}
    \centering
    \includegraphics[width=\columnwidth]{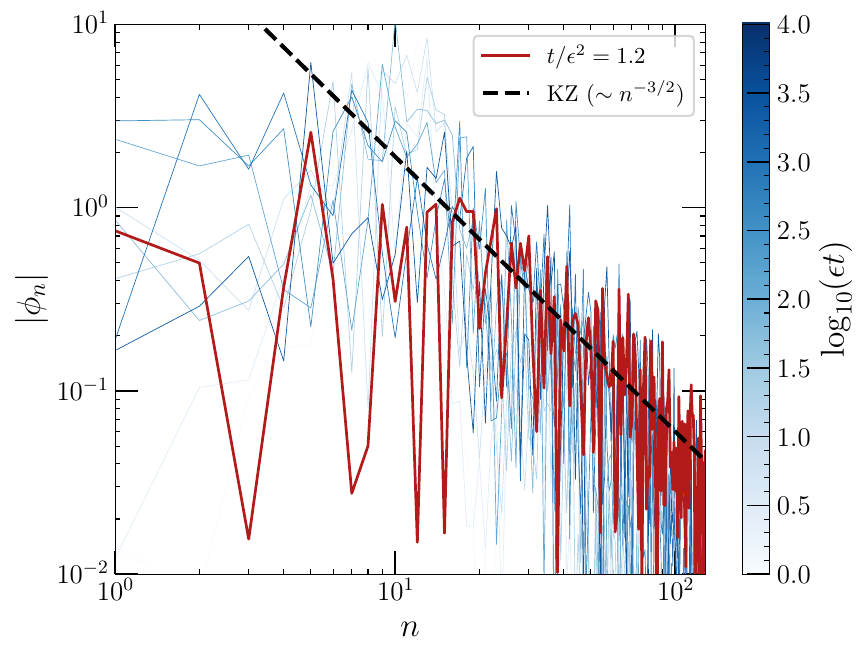}
    \caption{Field spectrum $|\phi_n|$ at different wave number $n$ for large initial data with amplitude $\epsilon = 20$, initially excited in the modes $n=10,11$. The different colors indicate the field spectrum at different times, as indicated by the color bar. The spectrum at the final time of the simulation, $t =5000 \approx \epsilon^2$ is shown as a wider, red line. It agrees with the predicted Kolmogorov-Zakharov scaling $n^{-3/2}$ (dashed black line). The resolution of this simulation is $N=1024$ modes.}
    \label{fig:1d_inertial}
\end{figure}

We prepare the evolution by exciting the two modes $n=10,11$, both with amplitude $\epsilon=20$ (i.e., far away from the expected regime of validity of perturbation theory), and evolve the system until (long times) $T \gtrsim \epsilon^{-2}$. In Fig.~\ref{fig:1d_inertial} we present the evolution of the resulting spectrum, where we can see the development of an inertial range, characterized by a polynomial law, which can be interpreted as a direct energy cascade toward higher frequencies. 

The development of an inertial range in the long-   term dynamics of this nonlinear equation suggests an interpretation in terms of wave turbulence~\cite{kolmogorov1968local, zakharov1965weak, zakharov2012kolmogorov,nazarenko2011wave}. We follow the derivation presented in~\cite{nazarenko2011wave}. Assuming that we treat small fluctuations with respect to the trivial solution, we can trade tracking the exact value of field modes $\{\phi_k(t)\}$ for the average particle number $n_k(t) \sim \langle |\phi_k(t)|^2\rangle$. The evolution equations for the field are replaced, then, by the kinetic wave equation 
\begin{equation}\label{kinetic_wave_equation}
    \frac{dn_k}{dt} = \mathcal{C}[n_k] \, , 
\end{equation}
where $\mathcal{C}[n_k]$ is the collision integral. We refer the interested reader to~\cite{nazarenko2011wave} for further details. A cubic nonlinearity leads to a so called four wave kinetic process. One way of seeing this is that, if we would draw the perturbative problem of the excitation of NLO modes from linearized solutions using Feynman diagrams, we would draw a vertex with four legs (three incoming linear modes can excite a NLO mode). Building upon this intuition, we can decompose the collision integral into two different contributions,  $\mathcal{C}[n_k] = \kappa_{2 \leftrightarrow  2}\mathcal{C}_{2 \leftrightarrow  2}[n_k] +\kappa_{3 \leftrightarrow  1} \mathcal{C}_{3\leftrightarrow 1}[n_k]$, where the $\kappa$ coefficients are numbers and of minor importance. Each of these is given by 
\begin{equation}\label{collision_integrals}
    \begin{aligned}
        \mathcal{C}_{2 \leftrightarrow  2}[n_k] =& \int dk_1 dk_2 dk_3 W_{2\leftrightarrow 2}  \Bigl[(n_1+n_2)n_3n_k\\
        &\hspace{3cm} -n_1n_2(n_3+n_k)\Bigr] \, , \\
        \mathcal{C}_{3 \leftrightarrow  1}[n_k] =& \int dk_1 dk_2 dk_3 W_{3\leftrightarrow 1}\Bigl[(n_1+n_2)n_3n_k\\
        &\hspace{3cm}-n_1n_2(n_3-n_k)\Bigr] \, , 
    \end{aligned}
\end{equation}
where the weights of the integrals are 
\begin{equation}
    \begin{aligned}
        W_{2\leftrightarrow 2} =& \delta(k_1+k_2+k_3-k)\delta(\omega_1+\omega_2-\omega_3-\omega_k) \, , \\
        W_{3\leftrightarrow 1} =& \delta(k_1+k_2+k_3-k)\delta(\omega_1+\omega_2+\omega_3-\omega_k) \, .
    \end{aligned}
\end{equation}
Recall that for the wave equation in the circle, the dispersion relation is just $\omega_k = |k|$. Thus, the $\mathcal{C}_{2 \leftrightarrow  2}$ operator corresponds to processes that conserve the particle number, while $\mathcal{C}_{3 \leftrightarrow  1}$ to processes that violate particle number conservation. Therefore, as opposed to other systems which also involve three wave interactions, such as the nonlinear Schr\"odinger equation~\cite{escobedo2015theory}, or a model for gravitational waves propagating in flat space~\cite{Galtier:2021ovg, Gay:2024kay, Gay:2025unv}, the system only has one conserved quantity (the energy presented in Eq.~\eqref{NL_Energy_Norm}). This implies that the only possible inertial regime in the turbulent state must be a direct cascade. On the other hand, if $\kappa_{3\leftrightarrow1}=0$, then there would be an additional conserved quantity (particle number), which would be associated to an inverse cascade. 

There are special solutions to the kinetic wave equation which are stationary, i.e., $\dot{n}_k = 0$, under the assumption that there is dissipation acting at low and high frequencies, and the system is being injected energy at some scale~\cite{kolmogorov1968local, zakharov1965weak} (which is not the case for the model we are interested). In this scenario, there exist solutions of the form $n_k \sim k^{-\nu}$, in the regime bounded by the dissipation scales, which is known as the inertial regime. These spectra are known as Kolmogorov-Zakharov (KZ) spectra, and there are many techniques devoted to compute the power of these spectra $\nu$ precisely. For four wave equations, the direct energy cascade is associated to a power $\nu = 3/2$~\cite{collot2024stability}. In Fig.~\ref{fig:1d_inertial} we are comparing the late-time solution with this prediction and found good agreement.

Equipped with these expectations, we can now address the question about the end state of the nonlinear wave equation in this two-dimensional model, for small initial data. If there is no dissipation, a direct energy cascade toward higher energies ensues, ultimately spreading energy across all possible modes. In the asymptotic state, $t\to\infty$ all modes are excited with negligible amplitude (since the initial amount of energy in the system was finite), and the solution loses regularity in the asymptotic time, as evidenced by the fact that the higher order norms diverge as $t\to\infty$. However, in most physical situations we expect the system to dissipate energy beyond some scale $k \geq k_{\rm Diss}$. The development of a KZ-like inertial range requires of continuous injection of energy, precisely to compensate this dissipation. In the absence of energy injection, the end state would just be $\Phi=0$, i.e., the trivial solution. We emphasize that other nonlinear equations (for example, those for which the global existence of solutions is not guaranteed) or initial data very far from the perturbative regime, where this analysis based on weak wave turbulence does not hold, might exhibit different phenomenology at late times, such as, e.g., time-periodic solutions. 

\subsection{Nonlinear waves on the sphere}

The previous model allowed for a straightforward analysis of the end state but has several limitations. One of the main limitations, in particular, is that the dispersion relation $\omega_k^2 = k^2$ only matches the dispersion relation of the long-lived modes in the eikonal regime. On the other hand, there is a physically motivated approximation that does not add much complexity, and can be put in closer contact with our original problem more directly. 

\begin{figure}
    \centering
    \includegraphics[width=\columnwidth]{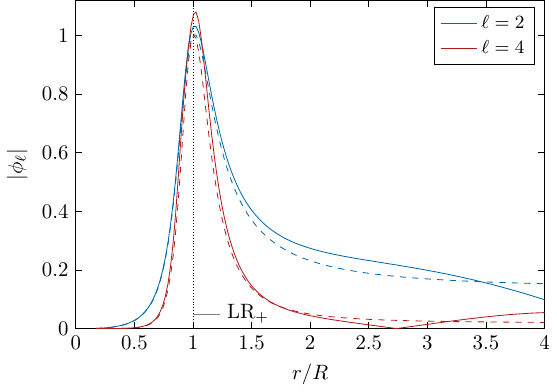}
    \caption{
    Normalized radial profile of the $\ell=2$ and $\ell=4$ modes extracted for a numerical simulation with weak coupling, $\epsilon = 0.1$, after $t/R=10^{3}$, where the initial data excites only the $\ell=2$ mode (solid lines). The dashed lines represent the radial profile of the fundamental mode associated to each $\ell$ mode. The agreement between both lines shows that most of the content of both linearly and nonlinearly excited modes, at large times, is contained in the fundamental mode, and peaks close to the stable LR.}
    \label{fig:Radial_Modes}
\end{figure}

After analyzing the radial structure of numerical solutions to the nonlinear wave equation on the model spacetime~\cite{Benomio:2024lev}, it is apparent that the waves are confined to a region close to the stable LR, i.e., at $r = R$. Indeed, Fig.~\ref{fig:Radial_Modes} shows the radial profile of the $\ell=2$ (the initially excited mode), and the $\ell=4$ mode (excited nonlinearly) at late times. The localization near the LR can be explained by comparing it with the radial profile of the fundamental modes, shown as dashed lines, which show good agreement (improved with increasing $\ell$ values). 

Accordingly, we can consider a simplified model, which consists of the dimensional reduction of our original problem to the sphere located at the LR (as done in Ref.~\cite{Benomio:2024lev} to measure the evolution of the $\ell$-spectrum). As we will show, eliminating the radial direction of the problem allows for significant simplifications. The radial dimension is important insofar as it provides the mechanism for dissipation through radiation leaking toward infinity. However, this dissipation can also be seen as a tunneling process, akin to quantum mechanics, without needing to resolve the dynamics in the radial direction. We can always add these tunneling rates a posteriori as dissipation coefficients, which are given by the damping times of the fundamental spherical harmonic modes. 

To be precise, let us focus at the surface $r=R$, where the induced metric is 
\begin{equation}
    ds_{r=R}^2 = -f(R)dt^2 + R^2d\Omega^2 = -dT^2 + R^2d\Omega^2 \, , 
\end{equation}
where we have rescaled time as $T = \sqrt{f(R)}t$. We examine the nonlinear wave equation in this geometry, i.e., $\Box_{g_R}\Phi = \kappa \Phi^3$, where now $\Phi:\mathcal{S}^2\times\mathbf{R}\to \mathbf{R}$ is a real field. Expanding in spherical harmonics, and assuming axisymmetry, we can write $\Phi = \sum_\ell \phi_\ell(T) \mathcal{Y}_{\ell,0}(\theta)$, so that the nonlinear wave equation reads
\begin{equation}
    \frac{d^2\phi_\ell}{dT^2}+\frac{\ell(\ell+1)}{R^2}\phi_\ell =-\kappa \sum_{\ell_1\ell_2\ell_3}\mathcal{I}_\ell^{\ell_1\ell_2\ell_3}\phi_{\ell_1}\phi_{\ell_2}\phi_{\ell_3} \, .
\end{equation}
Assuming that the initial data has small amplitude $\epsilon$, we can find perturbative solutions up to next to leading order. As in Sec.~\ref{sec:norms}, we write $\phi_\ell = \epsilon \phi^{(1)}_\ell + \epsilon^3 \phi^{(3)}_\ell$, and for initially static initial data, we have that the leading order solution is just 
\begin{equation}
    \phi_\ell^{(1)} = \Amp_\ell \cos(\omega_\ell T) \, , \qquad R\omega_\ell = \sqrt{\ell(\ell+1)} \, .
\end{equation}
Going back to the original time coordinate, we find that the oscillation frequency $\omega_\ell T = \nu_\ell t$, with $\nu_\ell^2 = \Omega \ell(\ell+1)$, matching the WKB result, which is also obtained when fitting the QNM frequencies obtained through direct integrations. Therefore, the reduction to the sphere captures accurately the real part of the frequency of the fundamental, long-lived modes. To NLO, writing $\Lambda = \{\ell \, \, | \, \, \Amp_\ell \neq 0\}$, we obtain the equation
\begin{equation}
    \ddot{\phi}^{(3)}_\ell + \frac{\ell(\ell+1)}{R^2}\phi^{(3)}_\ell = -\frac{\kappa}{4}\sum_{j\in\Lambda^3}\mathcal{I}_\ell^j \Amp_{j_1}\Amp_{j_2}\Amp_{j_3} \mathcal{F}_j(T) \, , 
\end{equation}
where $j=(j_1,j_2,j_3)$, and
\begin{equation}
    \mathcal{F}_j(T) = \sum_{\sigma_1,\sigma_2 = \pm1 }\cos(\Omega_j^{\sigma_1\sigma_2}T) \, ,
\end{equation}
with
\begin{equation}
    \Omega_j^{\sigma_1\sigma_2} = \omega_{j_1}+\sigma_1\omega_{j_2}+\sigma_2\omega_{j_3} \, .
\end{equation}
The $\ell$-th mode would be resonant if one of $\Omega^{\sigma_1\sigma_2}_j = \omega_\ell$ for some $j\in\Lambda^3$. If $j=\ell$, then $\Omega^{+-}_\ell = \Omega^{-+}_\ell = \omega_\ell$, which is actually the only possibility. In particular, if the initial data consists of a single mode, $\Lambda=\{\ell_0\}$, the only resonant mode at next to leading order is itself, $\ell_0$. For this case, we can then write the third order solution, as 
\begin{equation}
    \phi^{(3)}_\ell = \begin{cases}
    -\frac{\kappa \mathcal{I}_\ell^{\ell_0}\Amp_{\ell_0}^3}{4}\sum_{\sigma_1,\sigma_2=\pm1} \mathcal{G}_{\ell,\ell_0}^{\sigma_1,\sigma_2}(T) \, , & 
    \ell \neq \ell_0 \, , \\
    -\frac{\kappa \mathcal{I}_{\ell_0}^{\ell_0}\Amp_{\ell_0}^3}{8\omega_{\ell_0}}t\sin(\omega_{\ell_0}t) + \mathscr{O}(t^0)\, , & \ell = \ell_0 \, ,
    \end{cases}
\end{equation}
where 
\begin{equation}
   \mathcal{G}_{\ell,\ell_0}^{\sigma_1,\sigma_2}(T) = \frac{\cos(\omega_\ell T)-\cos(\Omega_{\ell_0}^{\sigma_1\sigma_2}T)}{\omega_\ell^2-(\Omega_{\ell_0}^{\sigma_1\sigma_2})^2} \, ,
\end{equation}
is a purely oscillatory term, which combines the homogeneous modes and the nonlinearly excited (cubic) modes. Note that we write $\Omega^{\sigma_1\sigma_2}_{\ell_0} = \Omega^{\sigma_1\sigma_2}_{(\ell_0,\ell_0,\ell_0)}$. Therefore, for $\ell \neq \ell_0$ (e.g., for $\ell = 3\ell_0$) the solution is purely oscillatory, and it excites both homogeneous and cubic modes with an amplitude which is cubic in the linear amplitude $\Amp_{\ell_0}$. The $\ell=\ell_0$ mode, however, features a secular linear growth in time, as well as subleading, oscillatory pieces. The results are the same for a general set of initially excited modes, $\tilde{\Lambda}$. Generically we have that $\phi_\ell^{(3)} \sim t \sin(\omega_\ell T)$ if $\ell \in \tilde{\Lambda}$, and otherwise, $\phi_\ell^{(3)} \sim \cos(\omega_\ell T)$ whenever $\ell \notin \tilde{\Lambda}$. This implies that all energy norms grow at the same rate, independently of the order of the norm. 

Although resonances only occur exactly within each mode separately, at large values of $\ell$ we have that $R \omega_\ell \approx \ell + 1/2$. Then, in the large $\ell$ limit, we have 
\begin{equation}
    \Omega_j^{+-} = \omega_{j_1+j_2-j_3} \, , \qquad \Omega_j^{++} = \omega_{j_1+j_2+j_3+1} \, ,
\end{equation}
and similarly for the other two possible combinations.
Therefore, as $\ell$ grows more modes become almost exactly resonant. This dispersion relation is also the dispersion relation of linear waves on the M\"obius open strip. Intuitively, this can be seen as arising from the fact that axisymmetric waves on the sphere are waves in the closed interval $[0,\pi]$, subject to the periodicity relation $\Phi(\theta) = \Phi(\theta+2\pi)$. A simple way to achieve this is by requiring the antiperiodicity condition $\Phi(\theta) = -\Phi(\theta+\pi)$, which under an appropriate rescaling is the condition naturally satisfied by sections of the fiber bundle associated to the M\"obius strip double cover of the circle. 

To understand the dynamics at longer timescales, we will solve the equations numerically. To facilitate a better comparison with the complete system, we will include the dissipation rate for each mode by formulating the evolution equation for each mode as follows:
\begin{equation}\label{2d_dissipation_eq}
    \ddot{\phi}_\ell + \frac{\ell(\ell+1)}{R^2}\phi_{\ell}+\frac{2}{\hat{\tau}_\ell}\dot{\phi}_{\ell} = -(\Phi)^3_{\ell} \, , 
\end{equation}
where the right hand side contains the nonlinear term, and we have added a dissipative term with a timescale $\hat{\tau}_\ell = \tau_{\ell}/(\Omega R^2)$, with $\tau_{\ell}$ the damping time obtained by fitting the WKB results to an exponential law. This equation is solved numerically using the same code as for the two dimensional scenario, but expanding instead on a basis of Legendre polynomials. We first test that, by doing this, we can reproduce the dynamics of the four-dimensional problem, at least in the regime where the initial data is small enough. 
\begin{figure}
    \centering
    \includegraphics[width=\columnwidth]{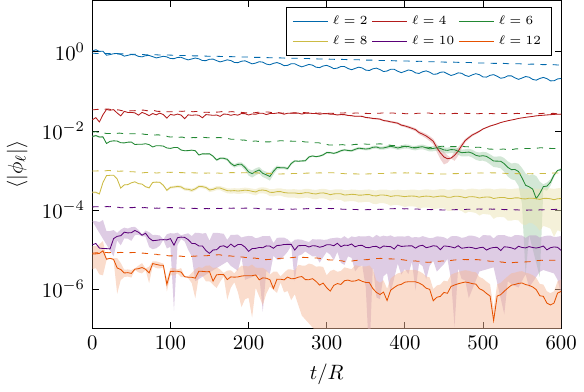}
    \caption{Root mean square value of the amplitude of each mode $\phi_\ell$ extracted at the LR for an evolution of the complete spacetime, with amplitude $\epsilon = 1/2$ in the notation of Ref.~\cite{Benomio:2024lev} (solid lines), and for a comparable amplitude, including artificial dissipation, for the nonlinear wave equation on the sphere (dashed lines). The shaded bands denoted an estimate of the numerical error for the $3+1$ evolution, obtained by comparing two different resolutions. The agreement is remarkable for the most dominant modes at early times, but it breaks down at later times, and for the higher $\ell$ modes.}
    \label{fig:Comparison_Weak}
\end{figure}

Figure~\ref{fig:Comparison_Weak} shows the numerical solution to the four-dimensional problem extracted at the LR as solid lines, and the numerical solution to~\eqref{2d_dissipation_eq} as dashed lines, after matching the amplitude of the initial data and rescaling the time appropriately. The uncertainty due to numerical error is shown as solid translucent bands, and is estimated by comparing two runs with resolution $\Delta x = 2.5 \times 10^{-3}$, and $\Delta x/2$. Clearly the numerical error is significant for the large $\ell$ modes after $t/R \gtrsim 100$. However, even the $\ell = 8, 10$ modes are reasonably well resolved up to the times shown in the Figure, and we can draw some conclusions by comparing with the $2+1$ model. 

The agreement for the lower $\ell$ modes is very good, especially at early times. At late times, even for low $\ell$, we observe oscillations which are not present in the $2+1$ model. These are likely due to the excitation of different overtones within each angular harmonic $\ell$, which have different amplitude at the LR.
We also observe that the agreement is much worse for the higher $\ell$ modes, even at early times. There is a physical reason why we should not expect the higher $\ell$ modes to agree as well as the lower $\ell$ modes to a dimensionally reduced problem: higher $\ell$ modes support several long-lived modes (overtones), where only the fundamental mode is confined at the LR.

\begin{figure}
    \centering
    \includegraphics[width=\columnwidth]{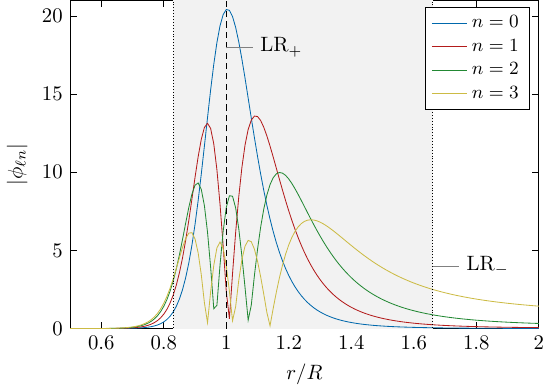}
    \caption{Normalized radial profile of the long-lived overtones of the $\ell = 10$ angular harmonic. Although the fundamental mode is localized very close to the LR (vertical-black dashed line), the higher overtones span a larger radial width, approximately ranging between the unstable LR, and its symmetric point in the potential, occupying all the filled gray region. }
    \label{fig:Radial_Confinement}
\end{figure}

On the one hand, the fundamental mode $n=0$ is always a long-lived mode, and has a radial support centered around the LR, with a radial spread $\delta^{(r)}_{\ell,n=0} \sim \ell^{-1/2}$~\cite{Cardoso:2014sna}. On the other hand, the highest overtone which is still long lived~\footnote{By this, we mean the highest overtone, such as its damping time being larger than a certain threshold, or simply the largest overtone number that can be found as a solution to the Bohr-Sommerfeld quantization condition in the WKB regime.}, $n=N_\ell$, has a radial support which does not depend on $\ell$, $\delta^{(r)}_{\ell,n=N_\ell} \sim r_{\rm LR_-}$. The modes with intermediate overtone number, for a fixed value of $\ell$, have a radial width which is proportional to its overtone number. Therefore, once these higher overtones start to play an important role in the nonlinear dynamics, the problem is not confined to a sphere located \emph{exactly} at the LR, but rather at a spherical shell with finite thickness, that supports a finite amount of modes for each angular harmonic $\ell$. We can see this phenomenon in Fig.~\ref{fig:Radial_Confinement}. 

This analysis implies that the dimensional reduction to the sphere might not be a good approximation for high-frequency modes since long-lived overtones may play a significant role. However, we can still gain some insights by analyzing the behavior at large times outside of the perturbative regime. In comparison to the two-dimensional model, we can anticipate a difference: In $1+1$ dimensions the dispersion relation was simply $\omega_k^2 = k^2$, hence, $\omega_{k_1+k_2+k_3}=\omega_{k_1}+\omega_{k_2}+\omega_{k_3}$, and as a consequence, the kinetic wave equation received contributions from the collision operator $\mathcal{C}_{3\leftrightarrow 1}$. However, in the sphere the dispersion relation is $R^2\omega_\ell^2 = \ell(\ell+1)$, and the combination of three modes onto a single mode is not resonant. 

\begin{figure}
    \centering
    \includegraphics[width=\columnwidth]{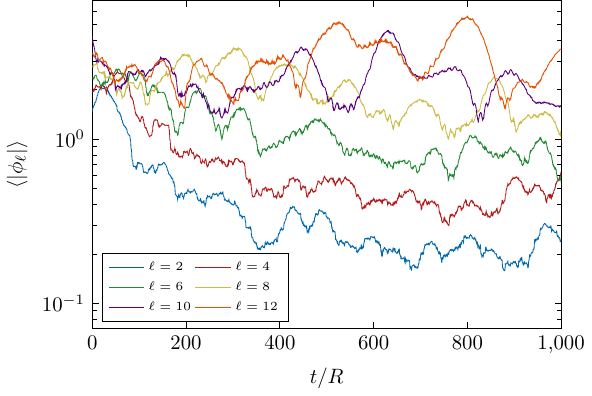}
    \caption{Evolution of the root mean square value of the mode amplitudes for some modes of the system, as depicted in the legend. The system is initialized in the $\ell=10,11$ modes, both with equal amplitude (total initial amplitude $\epsilon = 20$), evolved with a total resolution $N=256$. We observe hints of an inverse cascade, as a correlated growth/decay of the amplitudes of lower $\ell$ / higher $\ell$ modes. 
    \label{fig:2d_long_time}}
\end{figure}

We can also observe this behavior by analyzing the NLO in perturbation theory. In particular, since the kinetic wave equation does not receive contributions from $\mathcal{C}_{3\leftrightarrow 1}$, and this term was responsible for the violation of  the conservation of particle number, in its absence, we expect that the system undergoes both a direct energy cascade and an inverse cascade governed by the particle number (also known as \emph{wave action})~\cite{nazarenko2011wave,Gay:2025unv}. We can find hints of this process happening in Fig.~\ref{fig:2d_long_time}, e.g., in the form of ``revivals'' of low $\ell$ modes. These appear, for example, in the form of changes in the amplitudes of a lower $\ell$ mode and a decay of the amplitude of higher $\ell$ modes. For example, near $t/R\sim 450$ we see the amplitude of the $\ell=6,10$ modes grow significantly, while there is a decay in the amplitude of $\ell = 12$. Notice, for instance, the similarities between the dynamics of these modes, and the dynamics of radial modes of a scalar field in AdS for initial conditions within islands of stability~\cite{Balasubramanian:2014cja}, where also an inverse cascade is present~\cite{Buchel:2014xwa}. 

\begin{figure}
    \centering
    \includegraphics[width=\columnwidth]{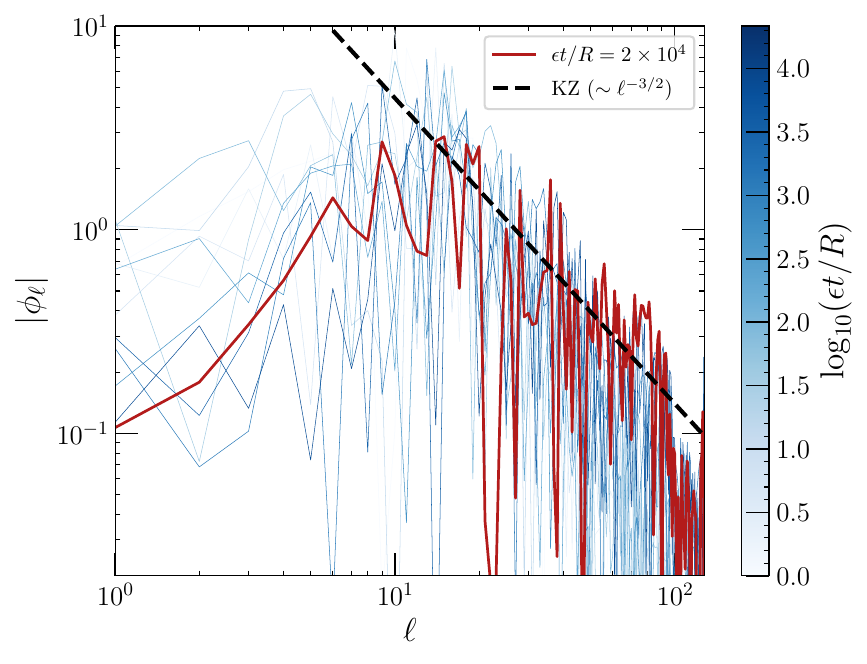}
    \caption{The power spectrum $|\phi_\ell|$ at different times (as indicated by the color bar), for the same configuration presented in Fig.~\ref{fig:2d_long_time}. The latest time available is represented with a thick-dark red line. At high frequencies, the spectrum is consistent with a Kolmogorov-Zakharov-type scaling $\phi_\ell \sim \ell^{-3/2}$ (black dashed line), as in the two dimensional model.}
    \label{fig:2d_inertial}
\end{figure}

The long-term dynamics of nonlinear waves on the sphere, in the presence of artificial dissipation in the manner of Eq.~\eqref{2d_dissipation_eq}, results of a combination of several effects. On the one hand, we have a dual cascade: energy cascade, from low to high frequencies, and particle number cascade, from high to low frequencies. However, we also have dissipation, which is (exponentially) more efficient at low frequencies. Due to the presence of this low-frequency dissipation, the inverse energy cascade just facilitates the dissipation of energy at low frequencies in the system. The resulting turbulent behavior is, thus, only dominated by the direct energy cascade, as can be seen in Fig.~\ref{fig:2d_inertial}. The Figure shows that the high frequency sector seems described by a KZ scaling. Although it is rather difficult to extract precise information about the KZ exponent, the scaling $\phi_\ell \sim \ell^{-3/2}$, characteristic of a direct energy cascade on a four wave interaction~\cite{Guo:2024cts}, appears as a good approximation. On the other hand, at low frequencies we observe the effect of dissipation, which suppresses significantly the spectrum for $\ell \lesssim 10$.

The development of an inertial range requires the dissipation of high-frequency radiation. However, the nonlinear wave model considered here lacks any intrinsic mechanism to provide this. In realistic scenarios, such dissipation can arise from the absorption of scalar waves or GWs by matter~~\cite{grishchuk1980gravitational, esposito1971absorption, papadopoulos1985absorption, Kocsis:2008aa, Kocsis:2011dr, Golat:2019aap, Boyanov:2024jge, Redondo-Yuste:2024vdb}, or from quantum effects.

For a KZ spectrum to persist in the inertial range, energy must be injected at a constant rate. One possible mechanism for this is accretion, which could inject GWs into the stable LR and sustain the KZ spectrum. However, this requires additional physical ingredients beyond the mere existence of a stable LR. In the absence of continuous energy injection, dissipation--inevitable at low frequencies, even in vacuum--dominates, and the KZ spectrum is not a stationary solution. Consequently, in the regime of validity of weak wave turbulence, we conjecture that the end state is always the trivial one, $\Phi = 0$.

\section{Discussion and Future Outlook}
\label{sec:discussion}

Several horizonless black hole mimickers and higher-dimensional spacetimes are known to admit stable light rings, which can efficiently trap high-frequency radiation, including gravitational waves. Consequently, small perturbations in such spacetimes decay slowly~\cite{Keir:2014oka, Benomio:2018ivy}. While it has been conjectured that this slow decay could trigger a nonlinear instability, the existing numerical and analytical evidence remains inconclusive.

Recent numerical work~\cite{Benomio:2024lev} has provided evidence that a stable LR can induce a turbulent cascade capable of transferring energy to higher frequencies. In this work, we extend this analysis by further investigating the dynamics of this cascade in the perturbative regime. Our results indicate that the sole presence of a stable LR does not necessarily lead to instability. In particular, for the nonlinear cubic wave equation considered here, we argue that the trivial solution $\Phi=0$ remains stable under small perturbations. 

One important aspect of our analysis concerns the regularity of solutions. Nonlinear wave equations are generally prone to energy-cascading phenomena. By examining a two-dimensional model, we have confirmed that higher-order norms of the solution grow rapidly~\cite{Bizon:2015pfa}, with a rate that increases with wave number. This growth leads to a gradual loss of smoothness over time. We find that a similar behavior occurs in the presence of a stable LR, reinforcing the evidence for a direct energy cascade. However, while such turbulence redistributes energy, it does not necessarily imply a global instability. In particular, the absence of an uncontrolled energy accumulation in any particular mode suggests that the system does not undergo a catastrophic breakdown.

A key question is whether this turbulent cascade could still lead to a physical instability over sufficiently long timescales. To address this, we have studied the long-term dynamics of a dimensionally reduced system, which describes waves propagating exactly at the LR. This model effectively captures the dynamics of low-$\ell$ modes, and our numerical simulations reveal hints of a dual cascade---where energy flows both toward higher frequencies (direct cascade) and toward lower frequencies (inverse cascade). When artificial dissipation is introduced at high frequencies, an inertial range emerges, characterized by a Kolmogorov-Zakharov spectrum. However, it is important to recognize that this model may not accurately describe the high-frequency behavior. In particular, a family of long-lived radial overtones exists for large angular numbers, supported within a finite-thickness spherical shell centered around the light rings. This suggests that conclusions from the reduced model should be treated cautiously, and further research is needed.

If the considered system with a stable light ring, were to be unstable, and the nonlinear backreaction of the scalar field on the geometry were included, at least the following possible outcomes could be expected: the formation of a black hole at the LR; the emergence of time-periodic solutions (e.g., quasimode solutions exhibiting an uniform decay rate slower than inverse logarithmically in time); or the evolution toward a state \emph{without} a stable LR altogether (as, e.g., conjectured in Ref.~\cite{Cunha:2022gde} for spinning boson stars but recently reconsidered in Ref.~\cite{Marks:2025jpt}). Regarding the first scenario, since radiation is only partially trapped within a finite-width shell, the formation of a black hole starting from arbitrarily small initial data appears inconceivable. This expectation aligns with the arguments presented in~\cite{Benomio:2024lev}. Unlike the case of pure AdS, where nonlinear wave interactions lead to the focusing of energy at a single point~\cite{Bizon:2011gg}, in this type of spacetimes, the energy remains distributed across an extended region. Additionally, we find no evidence for the formation of time-periodic solutions, aside from a turbulent Kolmogorov--Zakharov spectra. However, for these characteristic spectra to be maintained, both (i) high-frequency dissipation and (ii) a continuous energy injection mechanism are required. In the absence of sustained energy input, they must eventually decay. Our results, therefore, suggest that, after a sufficiently long time, the radiation either escapes the light ring or disperses into high-frequency modes with negligible amplitude. Based on this analysis, nonlinear waves propagating in spacetimes with stable light rings appear dynamically stable within the framework considered here. 

Several aspects remain unexplored in this work. If a black hole mimicker also possesses a horizon, in addition to a stable light ring, dissipation is enhanced, reinforcing the arguments against an instability. However, as noted in Ref.~\cite{Cunha:2017qtt}, topological considerations imply that in four-dimensional asymptotically flat spacetimes, the presence of a horizon and a stable light ring necessitates a second unstable light ring, introducing additional complexity. The interplay between multiple light rings warrants further investigation. Additionally, our analysis is based on a scalar field model, which, while capturing some essential nonlinear dynamics, does not account for full gravitational interactions. Since trapped radiation at the light ring can, in principle, modify the light ring's properties, the backreaction effects in a fully relativistic setting could introduce additional dynamical features that the nonlinear behavior arising solely from the self-interaction of the scalar field does not capture. Lastly, rotation can amplify radiation, either through partial absorption at the surface of the black hole mimicker or via the presence of ergoregions~\cite{Brito:2015oca}. Studying these effects in a well-posed matter model remains an important next step in confirming the conclusions drawn here.

\begin{acknowledgments}
We are very grateful to Frans Pretorius for inspiring the two-dimensional model we studied and for his feedback on all the stages of this work. We also thank Gabriele Benomio, Vitor Cardoso, and Nils Siemonsen for helpful discussions and comments.  
J.R-Y. is grateful to the Princeton Gravity Initiative for hospitality during the early stages of development of this work. 
J.R-Y. acknowledges support by VILLUM Foundation (grant no. VIL37766) and the DNRF Chair program (grant no. DNRF162) by the Danish National Research Foundation. The Center of Gravity is a Center of Excellence funded by the Danish National Research Foundation under grant No. 184.
A.C.-A. acknowledges support from the DOE through Los Alamos National Laboratory (LANL) Directed Research and Development, Grant No. 20240748PRD1, as well as by the Center for Nonlinear Studies. This work is authorized for unlimited release under LA-UR-25-21674. 
The Tycho supercomputer hosted at the SCIENCE HPC center at the University of Copenhagen was used for supporting this work.

\end{acknowledgments}

\appendix

\section{The WKB Approximation}
\label{App:WKB}

In this section we will briefly review the application of the Wentzel-Kramers-Brillouin (WKB) approximation to find the long-lived modes of a system. The approximation in this case corresponds to a perturbative expansion in $1/\ell$, and thus is accurate for high wave numbers. However, it has also been shown to be very accurate~\cite{Pani:2009ss, Cardoso:2014sna} in the presence of long-lived modes, i.e., in order to capture modes of the system with very large quality factors. 

Let us begin by considering the WKB ansatz, $\phi_\ell \sim e^{p S_p(r)-i\omega_p t}$, with $p = \ell+1/2$ so that the linear equation becomes 
\begin{equation}
    \frac{S^{\prime\prime}}{p} + (S^\prime)^2 + \Bigl(\Tilde{\omega}_p^2 - \mathcal{U}\Bigr) = \frac{\chi}{p^2} \, , 
\end{equation}
where the prime denotes derivatives with respect to the tortoise coordinate, i.e., $S^\prime = f \partial_r S$, the rescaled frequency is $\Tilde{\omega} = \omega/p$, and the potential splits into the WKB potential $\mathcal{U} = fr^{-2}$ and the correction $\chi$, given by 
\begin{equation}
    \chi = -\frac{f}{4r^2}+\frac{f^\prime}{r} \, .
\end{equation}
The solution between the turning points $r_a$ and $r_b$ can be found as an asymptotic series in $1/p$. Indeed, writing $S = S_0 + S_1/p + S_2/p^2$, one gets
\begin{equation}
    \begin{aligned}
        S_0 =& \pm i \int_{r_a}^r \sqrt{|\mathcal{Q}(x)|}\frac{dx}{f(x)} \, , \qquad
        S_1 = -\frac{1}{4}\log\mathcal{|Q|} \, \\ 
        S_2 =&  \pm \frac{i}{8} \int_{r_a}^r\Biggl[\frac{\mathcal{U}^{\prime\prime}(x)}{\mathcal{Q}(x)^{3/2}}+\frac{5}{4}\frac{\mathcal{U}^\prime(x)}{\mathcal{Q}(x)^{5/2}}-\frac{4\chi(x)}{\sqrt{\mathcal{Q}(x)}}\Biggr]\frac{dx}{f(x)} \, ,
    \end{aligned}
\end{equation}
where $\mathcal{Q}(x) = \Tilde{\omega}_p^2-\mathcal{U}(x)$. Remarkably, the first two terms in the expansion do \emph{not} depend on the source term $\chi$~\cite{Dias:2012tq}. Therefore, we will truncate the expansion at the first order in $p$. 

We now can use this expression to construct the solution in the four regions in which our problem is divided. We start from the outermost region, which we label region IV, where $r_c\leq r < \infty$. In general, the solution in that region, which is classically allowed, has the form 
\begin{equation}
    \phi_\ell^{IV} = \frac{\mathcal{A}}{\sqrt{p}\mathcal{Q}^{1/4}}e^{ip\eta_+(r,r_c)} + \frac{\mathcal{B}}{\sqrt{p}\mathcal{Q}^{1/4}}e^{-ip\eta_+(r,r_c)} \, , 
\end{equation}
where 
\begin{equation}
    \eta_{\pm}(r,r_a) = \int_{r_a}^r \sqrt{\pm\mathcal{Q}(x)} \frac{dx}{f(x)} \, ,
\end{equation}
are the WKB phases. 

As $r\to \infty$, $\mathcal{Q} \to \Tilde{\omega}$ becomes constant. Outgoing boundary conditions correspond to 
\begin{equation}
    \phi_\ell \to e^{i\omega r_\star} \, , 
\end{equation}
as $r \to \infty$. Thus, out-going boundary conditions are equivalent to requiring that $\mathcal{B} = 0$. For simplicity, we will then put $\mathcal{A}=1$, which is just an overall normalization constant. 

The solution in region III, $r_b\leq r \leq r_c$, can be written as 
\begin{equation}
    \phi_\ell^{III} = \frac{C_1}{\sqrt{p}|\mathcal{Q}|^{1/4}}e^{p\eta_-(r,r_b)} + \frac{C_2}{\sqrt{p}|\mathcal{Q}|^{1/4}}e^{-p\eta_-(r,r_b)} \, .
\end{equation}
The constants $C_1$ and $C_2$ are not independent, but instead are fixed by connection formulas, which match this solution to the solution in region IV. The connection formulas in this case yield 
\begin{equation}
    \begin{aligned}
        C_1 =& -ie^{-(p\eta_{cb} -i\pi/4)} \, , \\
        C_2 =& \frac{1}{2}e^{(p\eta_{cb} +i\pi/4)} \, ,
    \end{aligned}
\end{equation}
where $\eta_{cb}=\eta_-(r_c,r_b)$. Following the same logic, the solution in region II, $r_a\leq r\leq r_b$, has the form 
\begin{equation}
    \phi_\ell^{II} = \frac{C_3}{\sqrt{p}\mathcal{Q}^{1/4}}e^{i\eta(r,r_a)} + \frac{C_4}{\sqrt{p}\mathcal{Q}^{1/4}}e^{-i\eta(r,r_a)} \, ,
\end{equation}
where the connection formulas now yield 
\begin{equation}
    \begin{aligned}
        C_3 =& e^{-i(p\eta_{ba}-\pi/4)}\Bigl(C_2-\frac{i}{2}C_1\Bigr) \, , \\
        C_4 =& e^{i(p\eta_{ba}-\pi/4)}\Bigl(C_2+\frac{i}{2}C_1\Bigr) \, .
    \end{aligned}
\end{equation}
Finally, the solution in region I, $0<r\leq r_a$, is given by 
\begin{equation}
    \phi_\ell^I = \frac{C_5}{\sqrt{p}|\mathcal{Q}|^{1/4}}e^{p\eta_-(r, r_a)} + \frac{C_6}{\sqrt{p}|\mathcal{Q}|^{1/4}}e^{-p\eta_-(r, r_a)} \, ,
\end{equation}
with 
\begin{equation}
    \begin{aligned}
        C_5 =& -i\Bigl(C_3 e^{i\pi/4}+C_4e^{-i\pi/4}\Bigr) \, , \\
        C_6 =& \frac{1}{2}\Bigl(C_3 e^{i\pi/4}-C_4 e^{-i\pi/4} \Bigr)\, .
    \end{aligned}
\end{equation}
As $r \to 0$ the WKB phase becomes divergent, due to the divergence of the potential. Thus, in order to have a regular solution at $r \to 0$, we need $C_6 = 0$. Inserting the connection formulas the regularity condition becomes 
\begin{equation}
    e^{2ip\eta_{cb}}\Bigl(1+e^{2ip\eta_{ba}}\Bigr) + i\Bigl(e^{2ip\eta_{ba}}-1\Bigr) = 0 \, , 
\end{equation}
which, in the large $p$ limit is dominated by the first term, yielding
\begin{equation}
    e^{ip\eta_{ba}}+e^{-ip\eta_{ba}} = 0 \, .
\end{equation}
This condition is just the Bohr-Sommerfeld quantization rule
\begin{equation}
    p\eta_{ba} \equiv \int_{r_a}^{r_b} \sqrt{\omega^2-\mathcal{V}_\ell}\frac{dr}{f(r)} = \pi\Bigl(n+\frac{1}{2}\Bigr) \, ,
\end{equation}
where $n$ denotes the overtone index.

The damping time is related to the tunneling probability between regions II and IV, which scales as 
\begin{equation}
    \tau \sim e^{2\eta_{bc}} \, .
\end{equation}

We can now compute the leading order scaling of the frequencies with the angular number $\ell$. Indeed, by expanding around the stable light ring $r \sim R$, and keeping only the leading order term, $V \sim \ell^2 f(R)/R^2$, the Bohr-Sommerfeld condition yields
\begin{equation}
    \omega = \Omega \ell + \mathcal{O}(\ell^0) \, ,
\end{equation}
where $\Omega = \sqrt{f(R)}/R$ is the frequency of the stable LR. Similarly, for the imaginary part, we obtain 
\begin{equation}
    \log \tau \sim \gamma \ell + \mathcal{O}(\ell^0) \, , 
\end{equation}
where 
\begin{equation}
     \gamma = \frac{\pi (f_{\rm LR_-}-\Omega^2)}{\sqrt{2 Q^{(2)}_{\rm LR_-}}} \, , 
\end{equation}
with $f_{\rm LR_-}, Q^{(2)}_{\rm LR_-}$ as given in the main text.

\section{Breit-Wigner Resonances}
\label{App:BWResonances}

Chandrasekhar and Ferrari~\cite{Chandra:1991I, Chandra:1991III} noticed that compact stars close to the Buchdahl limit have long-lived modes. Usually we think of QNMs as poles of the scattering matrix, but an alternative viewpoint is to think them as Breit-Wigner resonances in the scattering amplitude. By using this idea, they designed an algorithm which is very efficient at finding the frequencies of these longest lived modes. 

\begin{figure}
    \centering
    \includegraphics[width=\columnwidth]{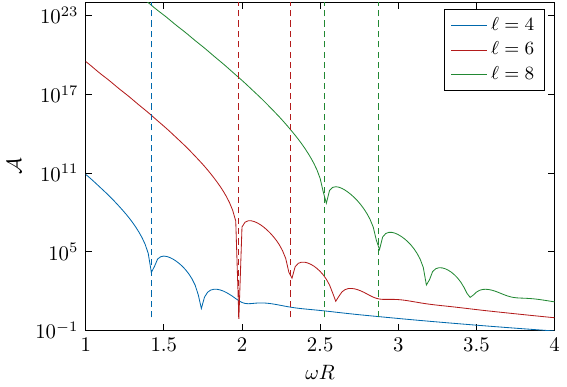}
    \caption{Breit-Wigner function $\mathcal{A}(\omega)$ for $\ell = 4,6,8$ and$\alpha = 0.166$. Each local minimum corresponds to a long-lived QNM. The dashed lines correspond to the calculation of the real part of the frequency using the WKB method, showing a good agreement between both methods. }
    \label{fig:Breit_Wigner}
\end{figure}

We start by integrate outward, ensuring that the solution is regular at the origin. At large distances, the outgoing solution takes the form $\phi_{\ell} \sim A \cos(\omega r) + B\sin(\omega r)$. Near the frequency of a long-lived mode, the scattering amplitude has the following behavior:
\begin{equation}
    \mathcal{A}(\omega) \equiv A^2+B^2 \sim (\omega-\omega_R)^2 + \omega_I^2 \, , 
\end{equation}
where the real part of the frequency $\omega_R$ acts as the effective energy of the state, and $\omega_I$ as its decay rate. Then, the longer lived modes can be found by identifying the minima of $\mathcal{A}(\omega)$, which correspond to $\omega_R$. Then, the imaginary part can be computed as 
\begin{equation}
    \omega_I = \sqrt{\frac{2\mathcal{A}(\omega_R)}{\mathcal{A}^{\prime\prime}(\omega_R)}} \, .
\end{equation}
It is important to note that this approach makes no approximations on the potential, as opposed to the WKB approach. Its main advantage with respect to the direct integration method, discussed below, is that in order to find the modes we just need to search for the minima of $\mathcal{A}$ along the real line, whereas in the direct integration case we search for zeros of the Wronskian in the complex plane, which is computationally more involved. The characteristic behavior of $\mathcal{A}(\omega)$, from where the real part of the QNM frequencies can easily be read, is shown in Fig.~\ref{fig:Breit_Wigner}. 

The Breit-Wigner method refines the frequency obtained from the WKB method. However, it requires significantly more numerical precision for higher angular numbers to accurately resolve the small imaginary part of the QNMs.

\section{Nonlinear Wave Equation in the Circle}
\label{App:NWE_Circle}

In this Appendix we provide additional details and formulas related to the perturbative solution to the nonlinear wave equation on the circle, as discussed in Sec.~\ref{sec:norms}. We remind the reader that the problem at hand is the initial value problem 
\begin{equation}
    \begin{cases}
        -\partial^2_t \Phi + \partial^2_x \Phi &= \Phi^3 \, , \\
        \Phi(t=0,x) &= \epsilon \cos(n x) \, , \quad \partial_t\Phi(t=0,x)=0 \, ,
    \end{cases}
\end{equation}
where $n$ is an integer number, and we look for a perturbative solution of the form $\Phi = \epsilon \phi^{(1)}+\epsilon^3\phi^{(3)} + \epsilon^5 \phi^{(5)}$. In the notation of the main text, we have 
\begin{equation}
    \begin{aligned}
        \phi^{(1)} =& \cos(nx) \, , \\
        \phi^{(3)}=& a_1(t)\cos(nx)+a_3(t)\cos(3nx) \, , \\
        \phi^{(5)} =& b_1(t)\cos(nx)+b_3(t)\cos(3nx)+b_5(t)\cos(5nx) \, ,
    \end{aligned}
\end{equation}
and the time dependent coefficients are given by 
\begin{equation}
    \begin{aligned}
        128n^2a_1 =& -3\Bigl[12nt\sin(nt)+\cos(nt)-\cos(3nt)\Bigr] \, , \\
        384n^2a_3 =& -9\Bigl[4nt\sin(3nt)+9\cos(nt)-9\cos(3nt)\Bigr] \,  ,\\
        (2^{14}3^2)n^4 b_1 =& (1814-5904n^2t^2)\cos(nt)-12nt\Bigl[\sin(5nt)\\
        &+249\sin(3nt)-708\sin(nt)\Bigr]\\
        &+103\cos(5nt)-1917\cos(3nt) \, , \\
        (2^{14}3^2)n^4 b_3 =& -(324+744n^2t^2)\cos(3nt) -4nt\Bigl[9\sin(5nt)\\
        &+19\sin(3nt)-261\sin(nt)\Bigr]\\
        &+99\cos(5nt)+225\cos(nt) \, , \\
        (2^{14} 15^2 )n^4 b_5 =& (25-360n^2t^2)\cos(5nt)-12nt\Bigl[129\sin(5nt)\\
        &-75\sin(3nt)+25\sin(nt)\Bigr] \, , \\
        &-1350\cos(3nt)+1325\cos(nt) \, .
    \end{aligned}
\end{equation}
Using these solutions, we can directly compute the Sobolev and high-order energy norms. We will first introduce some short-hand notation,
\begin{equation}
    \begin{aligned}
        \sigma_p =& \begin{cases} +1 \qquad & p \mod 4 = 0,3 \\ -1 & p \mod 4 = 1,2\end{cases} \, , \\
        \mathrm{CS}_p(x) =& \begin{cases}\cos(x) \qquad & \text{$p$ is even} \\ \sin(x) \qquad & \text{$p$ is odd} \end{cases} \, .
    \end{aligned}
\end{equation}
This is convenient, since $\partial_x^{(p)}\cos(x) = \sigma_p \mathrm{CS}_p(x)$. Now, let us begin by examining the Sobolev norms. Since we have that 
\begin{equation}
    \begin{aligned}
        (\partial_x^{(p)}\Phi)^2 =& \epsilon^2 (\partial_x^{(p)}\phi^{(1)})^2 + 2\epsilon^4 \partial_x^{(p)}\phi^{(1)}\partial_x^{(p)}\phi^{(3)} \\
        &+ \epsilon^6 \Bigl[(\partial_x^{(p)}\phi^{(3)})^2+2\partial_x^{(p)}\phi^{(1)}\partial_x^{(p)}\phi^{(5)}\Bigr]+\mathscr{O}(\epsilon^8) \, ,
    \end{aligned}
\end{equation}
we can directly compute each of the derivatives and integrate on the sphere. The result is
\begin{equation}
    \begin{aligned}
        \norm{D^{(k)}\Phi}^2 =& \epsilon^2 \pi n^{2k}\Bigl[\cos^2(nt) + 2\epsilon^2\cos(nt)a_1(t) \\
        &+\epsilon^4 \Bigl(2\cos(nt)b_1(t) + a_1(t)^2 + 3^{2k}a_3(t)^2\Bigr)\Bigr] \, .
    \end{aligned}
\end{equation}
The energy norms can be decomposed as 
\begin{equation}
    \begin{aligned}
        \mathbf{E}_{\rm nl}^{(k)}[\Phi](t) =& \norm{D^{(k)}\Phi}^2+\mathbf{T}^{(k)}[\Phi](t) + \mathbf{U}[\Phi](t) \, , \\
        \mathbf{T}^{(k)}[\Phi](t)\equiv& \int_{\mathbf{S}^1}dx (\partial_t^{(k)}\Phi)^2 \, , \qquad \mathbf{U}[\Phi](t)=\int_{\mathbf{S}^1}dx \frac{\Phi^4}{4} \, .
    \end{aligned}
\end{equation}
Therefore, we already have computed one of the pieces. The piece involving time derivatives leads to a more lengthy calculation, which leads to
\begin{equation}
    \begin{aligned}
        \mathbf{T}^{(k)}[\Phi](t) =& \pi \epsilon^2 \Biggl(n^{2k}\mathrm{CS}^2_k(nt) + 2\sigma_k n^k a^{(k)}_1\mathrm{CS}_k(nt)\\
        &+\Bigl[(a^{(k)}_1)^2+(a^{(k)}_3)^2+2\sigma_k n^k (b^{(k)})_1\mathrm{CS}_k(nt)\Bigr] \Biggr) \, .
    \end{aligned}
\end{equation}
Lastly, for the potential term, 
\begin{equation}
    \mathbf{U}[\Phi](t) = \frac{3\pi\epsilon^4}{8}\cos^4(nt)+\frac{\pi\epsilon^6}{2}(3a_1+a_3)\cos^3(nt) \, .
\end{equation}
Putting all the pieces together we have 
\begin{equation}
    \begin{aligned}
        \mathbf{E}&_{\rm nl}^{(k)}[\Phi](t) = \pi \epsilon^2 n^{2k} \Bigl[\cos^2(nt)+\mathrm{CS}_k^2(nt)\Bigr]\\
        &+2\pi\epsilon^4 n^k \Bigl[n^k a_1 \cos(nt) + \sigma_k a_1^{(k)}\mathrm{CS}_k(nt) +\frac{3}{16n^k}\cos^4(nt) \Bigr]\\
        &+\pi\epsilon^6\Bigl[(a_1^{(k)})^2+n^{2k}a_1^2+(a_3^{(k)})^2+(3n)^{2k}a_3^2\\
        &\quad +2n^k\Bigl(\sigma_k b_1^{(k)}\mathrm{CS}_k(nt) + n^k b_1 \cos(nt)\Bigr)\\
        &+\frac{1}{2}\Bigl((3a_1+a_3)\cos^3(nt)\Bigr)\Bigr] \, .
    \end{aligned}   
\end{equation}
When setting $k=1$ we find $ \mathbf{E}_{\rm nl}^{(1)}[\Phi](t) = \pi\epsilon^2 ( n^2+3\epsilon^2/8)$ for all times, therefore, it is just a constant. Notice how the potential term is necessary in order to cancel out all possible time dependence. This matches the energy norm evaluated directly on the initial data. 

\section{Convergence Test}
\label{Appe:ConvergeTests}

We have implemented two numerical codes to solve the nonlinear wave equation in the circle, and for axisymmetric modes in the sphere. As discussed in the main text, the implementation is based on a pseudospectral method. The Fourier coefficients of the field (in the circle), and the coefficients of the expansion on Legendre polynomials (on the sphere) are evolved in time using a fourth order Runge--Kutta method. We compute the nonlinear term in real domain, rather than in the spectral domain, using the \texttt{ApproxFun.jl} package to transform from real to Fourier (respectively, Legendre) domain, in every time step. 

\begin{figure}
    \centering
    \includegraphics[width=\columnwidth]{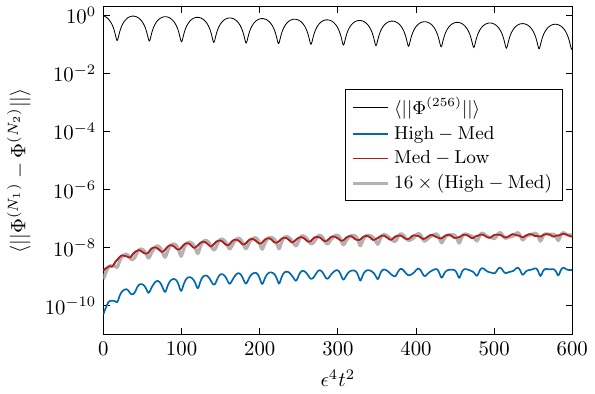}
    \caption{The Norm defined in Eq.~\eqref{Convergence_norm} for the highest resolution solution (using $N=256$) in solid, black line, compared with the residuals between medium and high resolutions ($N=128,256$, respectively) in blue, and the residual between medium and low resolutions ($N=128,64$, respectively), in red. Overlapped and as a thicker, translucent gray line, we represent the rescaled residual between the medium and high resolutions, scaled by the convergence factor consistent with fourth-order convergence.}
    \label{fig:2d_Convergence}
\end{figure}

An important test is to check the accuracy and convergence of our implementation. We do so by estimating the lowest order norm of the residual, defined as 
\begin{equation}\label{Convergence_norm}
    ||{\Phi^{(N_1)}-\Phi^{(N_2)}}|| \equiv \Biggl(\sum_{n\in \min(N_1,N_2)} |\phi^{(N_1)}_n-\phi^{(N_1)}_n|^2\Biggr)^{1/2} \, , 
\end{equation}
where $\Phi^{(N)}$ is the solution obtained using a total $N$ modes, and $\phi^N_n$ is the coefficient of each of the modes for that particular solution. In Fig.~\ref{fig:2d_Convergence} we show that the resolution error is several orders of magnitude below the norm of the solution, which estimates the accuracy of our method. Further, we show that doubling the resolution decreases the error by a factor of $2^4$, as expected for a fourth order method. Similar results are obtained also for the two-dimensional nonlinear wave equation in the circle, since the numerical implementation is identical.

\bibliography{biblio}

\end{document}